\documentclass[twocolumn, longauth]{aastex631}

\usepackage{savesym}
\savesymbol{tablenum}

\usepackage{graphicx}
\usepackage{txfonts}
\usepackage{hyperref}

\def \fluxdensityunitpix {{\rm erg\,s$^{-1}$\,cm$^{-2}$\,\AA$^{-1}$\,pix$^{-1}$}}

\def \fluxdensityunit {{\rm erg\,s$^{-1}$\,cm$^{-2}$\,\AA$^{-1}$}}

\newcommand{\Ha}{H{$\alpha$}}

\newcommand{\lea}{{\>\rlap{\raise2pt\hbox{$<$}}\lower3pt\hbox{$\sim$} \>}}
\newcommand{\gea}{{\>\rlap{\raise2pt\hbox{$>$}}\lower3pt\hbox{$\sim$} \>}}

\newcommand{\OSU}{\affiliation{Department of Astronomy, The Ohio State University, 140 West 18th Avenue, Columbus, Ohio 43210, USA}}

\newcommand{\Alberta}{\affiliation{Department of Physics, University of Alberta, Edmonton, AB T6G 2E1, Canada}}

\newcommand{\ANU}{\affiliation{Research School of Astronomy and Astrophysics, Australian National University, Canberra, ACT 2611, Australia}}

\newcommand{\IPARCOS}{\affiliation{Instituto de F\'{\i}sica de Part\'{\i}culas y del Cosmos, Universidad Complutense de Madrid, E-28040 Madrid, Spain}}

\newcommand{\Carnegie}{\affiliation{Observatories of the Carnegie Institution for Science, 813 Santa Barbara Street, Pasadena, CA 91101, USA}}

\newcommand{\CCAPP}{\affiliation{Center for Cosmology and Astroparticle Physics, 191 West Woodruff Avenue, Columbus, OH 43210, USA}}

\newcommand{\CfA}{\affiliation{Harvard-Smithsonian Center for Astrophysics, 60 Garden Street, Cambridge, MA 02138, USA}}

\newcommand{\ESO}{\affiliation{European Southern Observatory, Karl-Schwarzschild Stra{\ss}e 2, D-85748 Garching bei M\"{u}nchen, Germany}}

\newcommand{\Heidelberg}{\affiliation{Astronomisches Rechen-Institut, Zentrum f\"{u}r Astronomie der Universit\"{a}t Heidelberg, M\"{o}nchhofstra\ss e 12-14, D-69120 Heidelberg, Germany}}

\newcommand{\ICRAR}{\affiliation{International Centre for Radio Astronomy Research, University of Western Australia, 35 Stirling Highway, Crawley, WA 6009, Australia}}

\newcommand{\ITA}{\affiliation{Universit\"{a}t Heidelberg, Zentrum f\"{u}r Astronomie, Institut f\"{u}r Theoretische Astrophysik, Albert-Ueberle-Str 2, D-69120 Heidelberg, Germany}}

\newcommand{\IWR}{\affiliation{Universit\"{a}t Heidelberg, Interdisziplin\"{a}res Zentrum f\"{u}r Wissenschaftliches Rechnen, Im Neuenheimer Feld 205, D-69120 Heidelberg, Germany}}

\newcommand{\JHU}{\affiliation{Department of Physics and Astronomy, The Johns Hopkins University, Baltimore, MD 21218, USA}}

\newcommand{\MPE}{\affiliation{Max-Planck-Institut f\"{u}r extraterrestrische Physik, Giessenbachstra{\ss}e 1, D-85748 Garching, Germany}}

\newcommand{\MPIA}{\affiliation{Max-Planck-Institut f\"{u}r Astronomie, K\"{o}nigstuhl 17, D-69117, Heidelberg, Germany}}

\newcommand{\OAN}{\affiliation{Observatorio Astron\'{o}mico Nacional (IGN), C/Alfonso XII, 3, E-28014 Madrid, Spain}}

\newcommand{\OCAD}{\affiliation{OCAD University, Toronto, Ontario, M5T 1W1, Canada}}

\newcommand{\Ox}{\affiliation{Sub-department of Astrophysics, Department of Physics, University of Oxford, Keble Road, Oxford OX1 3RH, UK}}

\newcommand{\Princeton}{\affiliation{Department of Astrophysical Sciences, Princeton University, Princeton, NJ 08544 USA}}

\newcommand{\UToledo}{\affiliation{Ritter Astrophysical Research Center, University of Toledo, Toledo, OH, 43606}}

\newcommand{\UBonn}{\affiliation{Argelander-Institut f\"ur Astronomie, Universit\"at Bonn, Auf dem H\"ugel 71, 53121 Bonn, Germany}}

\newcommand{\UChile}{\affiliation{Departamento de Astronom\'{i}a, Universidad de Chile, Camino del Observatorio 1515, Las Condes, Santiago, Chile}}

\newcommand{\UCM}{\affiliation{Departamento de F\'{\i}sica de la Tierra y Astrof\'{\i}sica, Universidad Complutense de Madrid, E-28040 Madrid, Spain}}

\newcommand{\ULyon}{\affiliation{Univ Lyon, Univ Lyon 1, ENS de Lyon, CNRS, Centre de Recherche Astrophysique de Lyon UMR5574,\\ F-69230 Saint-Genis-Laval, France}}

\newcommand{\UniCA}{\affiliation{Université Côte d'Azur, Observatoire de la Côte d'Azur, CNRS, Laboratoire Lagrange, 06000, Nice, France}}

\newcommand{\UWyoming}{\affiliation{Department of Physics and Astronomy, University of Wyoming, Laramie, WY 82071, USA}}

\newcommand{\STScI}{\affiliation{Space Telescope Science Institute, 3700 San Martin Drive, Baltimore, MD 21218, USA}}

\newcommand{\INAF}{\affiliation{INAF -- Osservatorio Astrofisico di Arcetri, Largo E. Fermi 5, I-50157, Firenze, Italy}}

\newcommand{\Rad}{\affiliation{Elizabeth S. and Richard M. Cashin Fellow at the Radcliffe Institute for Advanced Studies at Harvard University, 10 Garden Street, Cambridge, MA 02138, U.S.A.}}

\begin{document}
\shorttitle{The PHANGS-HST-H$\alpha$ Survey}
\shortauthors{Chandar et al.}
\title{The PHANGS-HST-H$\alpha$ Survey: \\ Warm Ionized Gas Physics at High Angular resolution in Nearby GalaxieS with the Hubble Space Telescope}


\author[0000-0003-0085-4623]{Rupali~Chandar}\email{Rupali.Chandar@utoledo.edu}
\UToledo
\author[0000-0003-0410-4504]{Ashley~T.~Barnes}
\ESO
\author[0000-0002-8528-7340]{David~A.~Thilker}
\JHU
\author[0000-0002-2957-3924]{Miranda~Caputo}
\UToledo
\author[0009-0008-2929-6665]{Matthew~R.~Floyd}
\UToledo
\author[0000-0002-2545-1700]{Adam K. Leroy}
\OSU
\author[0000-0001-7130-2880]{Leonardo \'Ubeda}
\STScI
\author[0000-0002-2278-9407]{Janice C. Lee}
\STScI
\author[0000-0003-0946-6176]{Médéric Boquien}
\UniCA
\author[0000-0001-6038-9511]{Daniel Maschmann}
\STScI
\author[0000-0002-2545-5752]{Francesco Belfiore}
\INAF
\author[0000-0001-6551-3091]{Kathryn Kreckel}
\Heidelberg
\author[0000-0001-6708-1317]{Simon C.~O. Glover}
\ITA
\author[0000-0002-0560-3172]{Ralf S.\ Klessen}
\ITA
\IWR
\CfA
\Rad
\author[0000-0002-9768-0246]{Brent Groves}
\ICRAR
\ANU
\author[0000-0002-5782-9093]{Daniel A. Dale}
\UWyoming
\author[0000-0002-3933-7677]{Eva Schinnerer}
\MPIA
\author[0000-0002-6155-7166]{Eric Emsellem}
\ESO
\ULyon
\author[0000-0002-5204-2259]{Erik~Rosolowsky}
\Alberta
\author[0000-0003-0166-9745]{Frank Bigiel}
\UBonn
\author[0000-0003-4218-3944]{Guillermo Blanc}
\Carnegie
\UChile
\author[0000-0002-5635-5180]{M\'elanie Chevance}
\ITA
\author[0000-0002-8549-4083]{Enrico Congiu}
\UChile
\author[0000-0002-4755-118X]{Oleg V. Egorov}
\Heidelberg
\author[0000-0001-5310-467X]{Chris Faesi}
\MPIA
\author[0000-0002-3247-5321]{Kathryn Grasha}
\ANU
\author{Stephen Hannon}
\MPIA
\author[0000-0003-3917-6460]{Kirsten L. Larson}
\STScI
\author[0000-0002-1790-3148]{Laura A. Lopez}
\OSU \CCAPP
\author[0000-0001-7413-7534]{Angus Mok}
\OCAD
\author[0000-0002-3289-8914]{Justus Neumann}
\MPIA
\author[0000-0002-0509-9113]{Eve Ostriker}
\Princeton
\author[0000-0001-7876-1713]{Alessandro Razza}
\UChile
\author[0000-0003-0651-0098]{Patricia S\'anchez-Bl\'azquez}
\UCM
\IPARCOS
\author[0000-0002-6363-9851]{Francesco Santoro}
\MPIA
\author{Andreas Schruba}
\MPE
\author[0000-0003-0378-4667]{Jiayi~Sun}
\Princeton
\author[0000-0003-1242-505X]{Antonio Usero}
\OAN
\author[0000-0002-7365-5791]{E. Watkins}
\Heidelberg
\author[0000-0002-3784-7032]{Bradley C. Whitmore}
\STScI
\author[0000-0002-0012-2142]{Thomas G. Williams}
\Ox


\begin{abstract}
The PHANGS project is assembling a comprehensive, multi-wavelength dataset of nearby ($\sim$5-20~Mpc), massive star-forming galaxies to enable multi-phase, multi-scale investigations into the processes that drive star formation and galaxy evolution. To date, large survey programs have provided molecular gas (CO) cubes with ALMA, optical IFU spectroscopy with VLT/MUSE, high-resolution NUV--optical imaging in five broad-band filters with HST, and infrared imaging in NIRCAM$+$MIRI filters with JWST.  Here, we present PHANGS-HST-H$\alpha$, which has obtained high-resolution ($\sim$\,2 -- 10\,pc), narrow-band imaging in the F658N or F657N filters with the HST/WFC3 camera of the warm ionized gas in the first 19 nearby galaxies observed in common by all four of the PHANGS large programs. We summarize our data reduction process, with a detailed discussion of the production of flux-calibrated, Milky Way extinction corrected, continuum-subtracted H$\alpha$ maps. PHANGS-MUSE IFU spectroscopy data are used to background subtract the HST-H$\alpha$ maps, and to determine the [NII] correction factors for each galaxy. We describe our public data products\footnote{The data released as part of this work includes the reduced drizzled narrow-band images and the flux-calibrated, continuum-subtracted H$\alpha$ maps for each galaxy.  These images are available for download via MAST at \url{https://archive.stsci.edu/hlsp/phangs.html}, as well as at the Canadian Astronomy Data Centre as part of the PHANGS archive\ at \url{https://www.canfar.net/storage/vault/list/phangs/RELEASES}.} and highlight a few key science cases enabled by the PHANGS-HST-H$\alpha$ observations.  
\end{abstract}


\section{Introduction}
\label{sec:intro}

The warm ionized gas within galaxies plays a pivotal role in shaping their morphology and evolution.  
Discrete structures of ionized gas
include HII regions/complexes, planetary nebulae (PNe), and supernova remnants (SNRs), among others.  
In addition to H$\alpha$ emission originating from HII regions, some ionizing photons escape and can travel far from their birth sites into lower density material.  This diffuse ionized gas (DIG) can contribute a significant fraction of the total H$\alpha$ emission from spiral galaxies and offers insights into interstellar radiation transport and its broader environmental impact. Recent advances, notably through instruments like the Multi-Unit Spectroscopic Explorer (MUSE), have revolutionized our ability to study the distribution and properties of the low surface brightness DIG \citep[e.g.,][]{Walterbos94,Belfiore2022}, 
shedding light on its connection to stellar feedback processes.

The evolution in the \textit{size} of HII regions is critical as a diagnostic of stellar feedback, which is viewed as one of the most important but uncertain aspects of galaxy evolution.
Feedback from massive stars (e.g.,\ winds, radiation, supernovae) in clusters and associations exerts pressure on the surrounding natal gas and dust, initiating a dynamic evolution from classic Strömgren spheres (on scales of order a parsec) to larger bubbles 
and even the most massive super-bubble structures, such as the $\sim$kpc-size Phantom Void \citep{Barnes2023}.  The dominant feedback mechanism changes as a function of time and spatial scale \citep[e.g.,][]{rahner2017,chevance2020,olivier21}, with radiation pressure playing an important role early on.  As ionizing photons penetrate the surrounding material, they create over-pressured expanding regions of ionized gas, characterized by bubbles, perforated walls, and partial shells.  The morphology of these structures, including their radius, degree of perforation/irregularity, and hierarchical organization, reflects the complex interplay between stellar feedback and the surrounding interstellar medium (ISM) \cite[e.g.,][]{Haffner09}.

For galaxies at distances of $\approx5-20$~Mpc, it is challenging to identify and study individual HII regions, since ground-based resolution is typically limited to $\approx1\arcsec$ ($\approx50 - 100$~pc; \citealp{McLeod20,Barnes2021}) and compact HII regions observed in the Milky Way have scales of just a few parsecs or below \citep{Caswell1987,Churchwell2002,Anderson14}. By obtaining more detailed maps of H$\alpha$ emission at these spatial scales, we will be able to connect individual stellar sources of ionization to the ionized gas and study the structure of feedback-dominated regions in detail, including shells, partial shells driven by HII regions, or larger bubbles propelled by supernova contributions, as well as assessing the sources of previously identified faint emission features, such as diffuse ionized gas.  In all, new high-resolution observations will enable us to close the loop on the impact of stellar feedback on galaxy-wide evolution from the stellar sources to their influence on the surrounding ISM, the formation and evolution of structures such as bubbles, and their influence on the subsequent generation(s) of stars.  This is now particularly relevant in the era of sub-arcsecond resolution ($\sim $HST matched) studies of the ISM offered by the James Webb Space Telescope (JWST).

The Cycle 30 HST Treasury project, PHANGS-HST-H$\alpha$, provides high-resolution maps of H$\alpha$ emission across 19 nearby spiral galaxies observed as part of the Physics at High Angular Resolution in Nearby Galaxies (PHANGS) project, and is the subject of this work.  The multi-wavelength dataset includes high-resolution CO(2-1) maps from the Atacama Large Millimeter/submillimeter Array (ALMA) \citep{phangs-alma}, broad-band imaging in 14 filters (including the H$\alpha$ presented here) from the near-ultraviolet through the mid-infrared with HST \citep{Lee2022} and JWST \citep{Lee2023,Williams24}, and optical IFU spectroscopy with the VLT/MUSE \citep{Emsellem2022}.  

\begin{figure*}[!htp]
    \centering
\includegraphics[width=\textwidth]{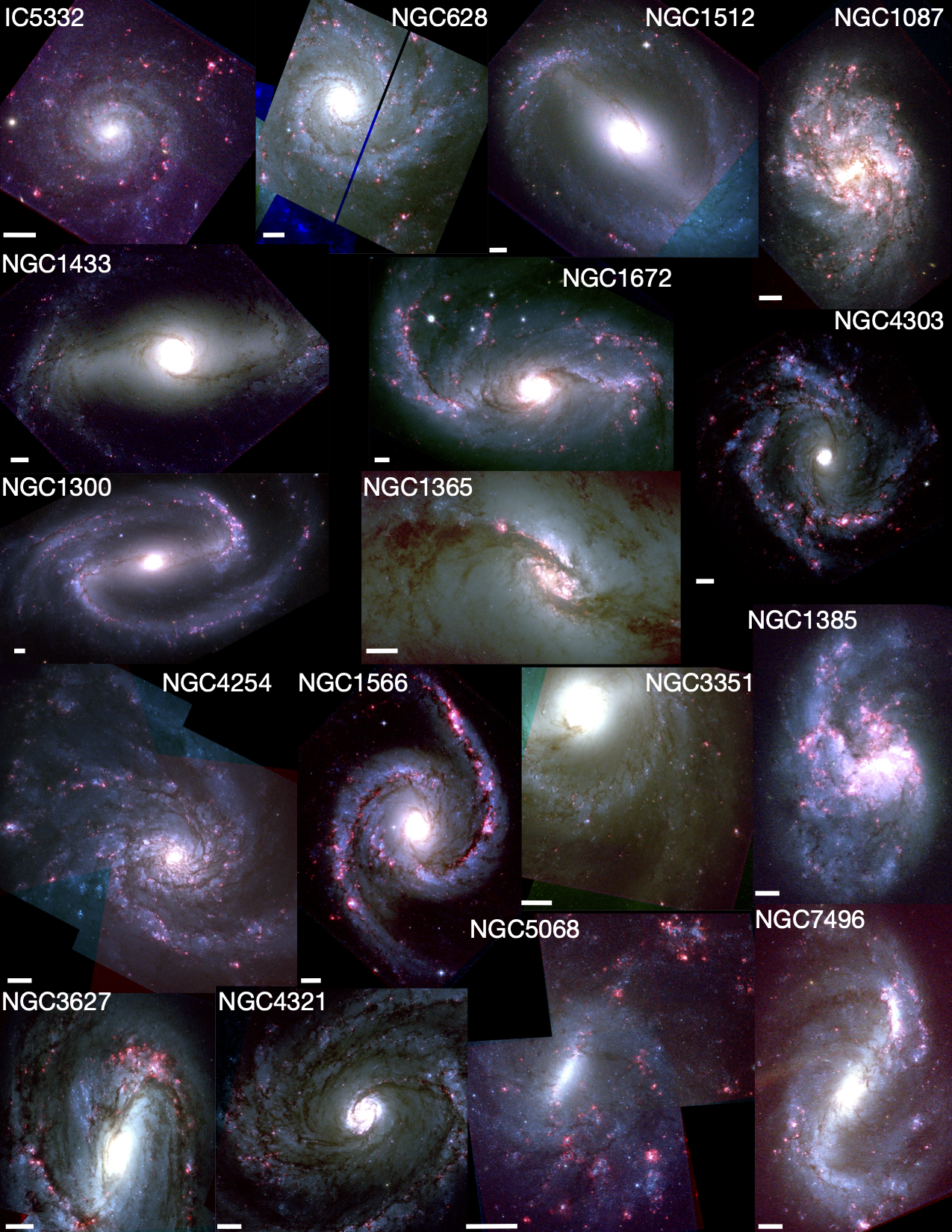}
    \caption{B-V-I+H$\alpha$ blue-green-red color composite images for the 17 (of 19) galaxies in our sample which have HST-H$\alpha$ images available at this time.  The H$\alpha$ emission (red) shows the locations of recent ($\lea 6$~Myr) star formation. A 1~kpc scale bar is indicated for each galaxy. The images show a variety of environments and morphologies.  
    \label{fig:pressrelease} } 
\end{figure*}

In this paper, we present the design of the PHANGS-HST-H$\alpha$ survey, an overview of the data processing pipeline, and describe the data products made public to the community.  The remainder of this paper is organized as follows.  In Section~\ref{sec:sample} we present the basic properties of our sample galaxies.  Section~\ref{sec:observations} summarizes the new HST data and reduction, as well as key characteristics of the MUSE/IFU spectra, which are used to improve the flux calibration of the HST images. Next, we explain in detail the flux calibration, continuum-subtraction, and correction for [NII] emission of the PHANGS-HST-H$\alpha$ images (Section~\ref{sec_fluxan}; see flowchart in Fig.\,\ref{fig_flowchart}).
We make the \texttt{PyHSTHAContSub} \citep{ashley_barnes_2025_14610187} pipeline for producing the flux calibrated continuum subtracted images publicly available online.\footnote{ See version 1 release of the code used throughout this work in the following link \url{https://zenodo.org/records/14610187}.}
Section~\ref{sec:dataproducts} presents the reduced data products and highlights key science areas where the H$\alpha$ maps will provide critical input. We summarize the key points of our work in Section~\ref{sec:summary}, including information on upcoming data products.

\section{Galaxy Sample}
\label{sec:sample}

The PHANGS multi-wavelength surveys focus on nearby, massive, and relatively face-on spiral galaxies. The parent sample (74 galaxies), described in \citet{phangs-alma}, selected spirals visible from the southern hemisphere (i.e., observable by ALMA and the VLT) and within $\approx20$~Mpc that have an inclination less than $75^{\circ}$.  The distance selection ensures that HST and JWST can resolve individual star clusters and associations, that MUSE can identify HII regions, and that ALMA resolves the ISM into molecular clouds and cloud complexes.  In order to focus on galaxies on the star-forming `main sequence', the environments where most massive stars form at $z=0$  \citep[][]{brinchmann04,salim07,Genzel10,Saintonge17}, the survey also selected galaxies with a stellar mass higher than $5\times10^9~M_{\odot}$ and a specific star formation rate (SFR$/M_{*}) \geq 10^{-11} {\rm yr}^{-1}$.  

Here, we present an overview of the PHANGS-HST-H$\alpha$ Treasury survey.  This project targets the 19 galaxies observed as part of the PHANGS-JWST Treasury program (PI: J. Lee); NGC~628, NGC~4254, and NGC~5068 have two pointings for a total of 22 fields.  All target galaxies have high-resolution imaging from HST available in five UV through optical broad-bands (PHANGS-HST; PI: J. Lee), millimeter-wave spectral mapping from PHANGS–ALMA \citep{phangs-alma}, and ground-based optical IFU spectroscopy from VLT/MUSE \citep{Emsellem2022}, as well as many other multi-wavelength observations.
The basic properties of the galaxies, including distance, the SFR (total and that in the area covered by the HST observations), and stellar mass are listed in Table~\ref{tab:galaxyprops}. Figure~\ref{fig:pressrelease} shows BVI$+$H$\alpha$ color images for the 17 out of 19 target galaxies that currently have H$\alpha$ images available. Young, recently formed stars appear blue; the pinkish-red emission shows the warm ionized gas that is the focus of this work. 
\begin{table*}[!htbp]
\caption{Properties of PHANGS-HST-H$\alpha$ Galaxies}
\centering
\begin{tabular}{llllllll}
\hline
\hline

\textbf{Galaxy} & $\alpha$ & $\delta$  &  Distance  & SFR$_{\rm tot}$ & SFR & log M$_{*}$  & log$_{10}L_{\rm CO}$\\ 
& [J2000] & [J2000] & [Mpc] & [M$_{\odot}$ yr$^{-1}$] &  [M$_{\odot}$ yr$^{-1}$] & [log M$_{\odot}$]  & [K~km~s$^{-1}$~pc$^{2}$]\\ 
(1) & (2) & (3) & (4) & (5) & (6) & (7) & (8)  \\ 
\hline
IC~5332    & 23h34m26.69s & -36d06m02.4s        & $9.01\pm0.41$             		&	0.41	& 0.11	&	9.5  & 7.1 \\ 
NGC~628-C$^*$     & 01h36m39.82s & +15d46m34.6s         & $9.84\pm0.63$             	&		1.75	& 0.93	&	10.3 & 8.4 \\ 
NGC~628-E     & 01h36m41.75s & +15d47m01.2s         & $9.84\pm0.63$             	&		1.75	& 0.93	&	10.3   &  8.4 \\ 
NGC~1087    & 02h46m25.09s & -00d29m54.8s       & $15.85\pm2.24$             	&		1.31	& 1.25	&	9.9  & 8.3 \\ 
NGC~1300$^*$    & 03h19m41.08s & -19d24m40.9s        & $18.99\pm2.85$ & 1.17 & 1.02 & 10.6  & 8.5 \\
NGC~1365    & 03h33m37.45s & -36d08m35.3s        & $19.57\pm0.78$             &		16.90	&12.88 	&	11.0 & 9.5 \\ 
NGC~1385    & 03h37m28.59s & -24d30m02.2s        & $17.22\pm2.58$             	&		2.09	&1.96	&	10.0 & 8.4 \\ 
NGC~1433$^*$    & 03h42m01.55s & -47d13m19.5s         & $18.63\pm1.86$    &	1.13	& 0.56	&	10.9 & 8.5 \\ 
NGC~1512    & 04h03m55.40s & -43d20m44.1s         & $18.83\pm1.88$             	&		1.28	& 0.59	&	10.7 & 8.3 \\
NGC~1566    & 04h20m00.64s & -54d56m14.6s     & $17.69\pm2.00$             	&		4.54	&3.21	&	10.8 & 8.9 \\ 
*NGC~1672$^*$    & 04h45m42.50s & -59d14m49.9s        & $19.40\pm2.91$             	&		7.60	&6.60	&	10.7  & 9.1  \\ 
NGC~2835-S    & 09h17m53.04s  & -22d23m01.2s        & $12.22\pm0.94$            	&	1.24	&0.57	&	10.0  & 7.7	\\ 
NGC~2835-N$^{\dagger}$    & 09h17m53.11s  & -22d21m19.5s          & $12.22\pm0.94$            	&	1.24	&0.57	&	10.0  & 7.7	\\ 
NGC~3351    & 10h43m57.95s & +11d42m18.7s        & $9.96\pm0.33$             	&		1.32	&0.87	&	10.4 & 8.1 \\ 
NGC~3627    & 11h20m15.53s & +12d59m59.3s & $11.32\pm0.48$             	&		3.84	& 3.31	&	10.8  & 9.0 \\ 
NGC~4254-E    & 12h18m56.31s & +14d26m03.5s        & $13.1\pm2.8$              	&		3.07	&2.77	&	10.4  &  8.9 \\ 
NGC~4254-W    & 12h18m46.95s & +14d24m36.9s        & $13.1\pm2.8$              	&		3.07	&2.77	&	10.4 & 8.9 \\ 
NGC~4303   & 12h21m55.40s & +04d28m29.7s        & $16.99\pm3.04$             &		5.33	&4.25	& 	10.5  & 9.0 \\ 
NGC~4321    & 12h22m54.57s & +15d49m19.4s         & $15.21\pm0.49$             &		3.56	&2.43	&	10.8 & 9.0 \\ 
NGC~4535   & 12h34m20.36s & +08d11m52.1s        & $15.77\pm0.37$             	&		2.16	&1.23	&	10.5  & 8.6 \\ 
NGC~5068-N    & 13h18m49.52s & -21d01m24.8s        & $5.20\pm0.21$             	&		0.28	&0.20	& 	9.4  & 7.3 \\ 
NGC~5068-S    & 13h18m53.92s & -21d03m10.4s        & $5.20\pm0.21$             	&		0.28	&0.20	&	9.4  & 7.3 \\ 
NGC~7496    & 23h09m47.36s & -43d25m31.4s        & $18.72\pm2.81$             	&		2.26	&2.00	&	10.0	& 8.3 \\ 
\hline
\hline
\end{tabular}
\tablecomments{Column~1: Galaxy Name. $^*$ indicates the HST H$\alpha$ observations are archival; $^{\dagger}$ indicates observations. NGC\,2835-N was observed just prior to publication, and the images for this galaxy will be included in the next image release.  Columns~2-3: R.A. and Dec.  Column~4: Distance in Mpc.
Columns~5-6: Star formation rate for the entire galaxy and the portion in the HST footprint (footprints are shown in the Appendix).
Column~7: Galaxy stellar mass.
Column~8: Integrated CO (2-1) luminosity.  Scale this by $\alpha_{\rm CO} \approx6.7M_{\odot}~\mbox{pc}^{-2}$ (K~km~s$^{-1}$)$^{-1}$ to estimate the molecular mass.
}
\label{tab:galaxyprops}
\end{table*}

Distance determinations to the galaxies in the PHANGS-HST-H$\alpha$ sample are taken from the compilation in \citet{Anand21}, and are based on a variety of methods. \citet{Anand21} determined new distances to 6 of these galaxies based on the tip of the red giant branch (TRGB). For the rest, they carefully assessed distance estimates available in the literature, which can vary in quality, and assigned an associated uncertainty.  Preference was given to high-quality primary indicators such as TRGB or Cepheid-based measurements, followed by estimates from other standard candles such as the planetary nebula luminosity function or surface brightness fluctuations. The lowest priority was given to distances based on modeling of local dynamical flows \citep{CosmicFlows1}.

Two estimates of the star formation rate (SFR) are listed in Table~\ref{tab:galaxyprops}. Column~5 compiles the total (SFR$_{\rm tot}$) which is estimated for the entire galaxy from the far ultraviolet luminosity measured from GALEX and the infrared luminosity measured from WISE W4, and follows the prescription given in \citet{leroy19} which is calibrated to match the results of the population synthesis modeling given in \citet{salim16,salim18}.  Column~6 provides the SFR, computed as for the entire galaxy, but limited to the area covered by the PHANGS-HST footprint (shown in the Appendix), which is similar to that covered by the HST-H$\alpha$ observations presented here.
The stellar mass estimates are from \citet{Leroy21} based on IRAC 3.6~$\mu$m observations when available, and WISE1 3.4~$\mu$m maps otherwise.  There is significant overlap in the wavelength coverage of the two bands, and both should be dominated by the light from old stars with little impact from dust extinction.  The near-infrared luminosity is combined with radially varying mass-to-light ratios, which are determined from constraints on the stellar population in each galaxy, to calculate the total stellar mass.  These values are compiled in Column~7 of Table~\ref{tab:galaxyprops}.
In Column~8 we compile the integrated CO (2-1) luminosity for each galaxy from \citep{Leroy2023}. The total mass of molecular gas $M_{\rm mol}$ can be estimated by assuming a fixed CO-to-H$_2$ conversion factor of  $\alpha_{\rm CO} \approx6.7M_{\odot}~\mbox{pc}^{-2}$ (K~km~s$^{-1}$)$^{-1}$.

\begin{figure*}[!htp]
    \centering
\includegraphics[width=\textwidth]{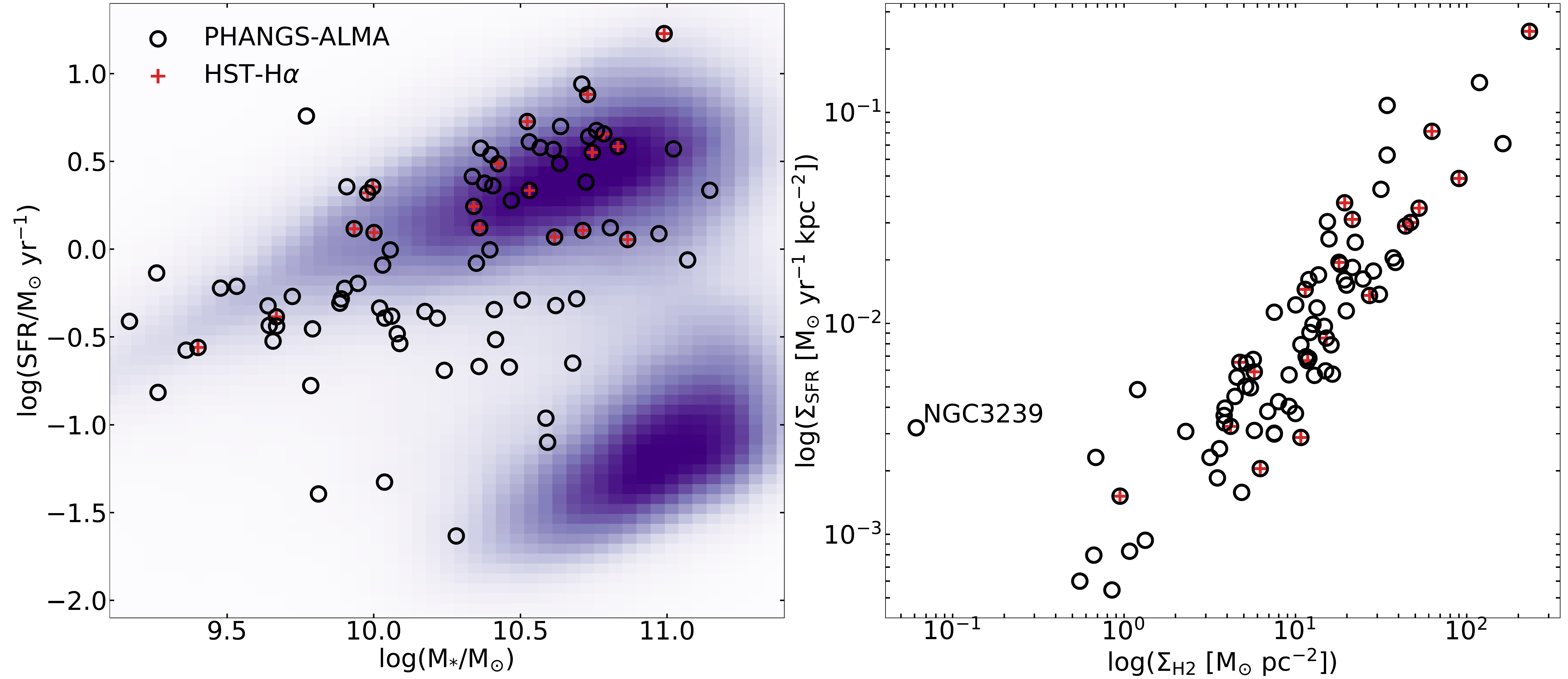}
    \caption{{\bf Left:} Galaxies from the SDSS sample with $z \lesssim 0.1$ form the backdrop of the left panel, where the galaxy star-forming sequence is the prominent upper feature. Coverage of the SFR--M$_*$ plane by galaxies in the parent PHANGS sample ($N=74$; open circles) is shown.
        Red crosses show that the 19 galaxies observed as part of the HST-H$\alpha$ project tend to have higher SFRs. 
        {\bf Right:} The star formation and molecular gas surface brightness \citep{Sun22} for the parent PHANGS sample (open circles) and the PHANGS-HST-H$\alpha$ subsample studied here (red crosses).  Outlier NGC~3239 is noted.        
    \label{fig:sample}}
\end{figure*}

Figure~\ref{fig:sample} shows basic properties of the PHANGS-HST-H$\alpha$ sample relative to the total PHANGS sample and to nearby star-forming galaxies in general. The left panel in Figure~\ref{fig:sample} shows $M_*$ vs. SFR for galaxies in the larger PHANGS sample and in the PHANGS-HST-H$\alpha$ subsample, overlaid on the locus occupied by star-forming galaxies from the Sloan Digital Sky Survey \citep{salim16}. 
The PHANGS samples provide excellent coverage of the main sequence located between stellar masses of ${\sim}10^{9.5} {-} 10^{11}~M_{\odot}$.  Main sequence galaxies in this stellar mass range are representative of the galaxies where the bulk of present-day star formation is taking place \citep{salim07, Saintonge17}.  The HST-H$\alpha$ subsample is weighted towards galaxies with somewhat higher SFR. Ongoing HST observations (see below, Sec.~\ref{sec:HSTdata}) will provide H$\alpha$ and NUV-U-B-V-I HST imaging for a combined total of 74 PHANGS galaxies, removing this bias.  The right panel of Figure~\ref{fig:sample} shows $\Sigma_{\rm SFR}$ (SFR per area) as a function of the molecular gas surface density for the full PHANGS sample (open circles) and for the PHANGS-HST-H$\alpha$ subsample (red crosses).

\section{HST and MUSE Observations}
\label{sec:observations}

\subsection{HST Data and Reduction}\label{sec:HSTdata}


Imaging observations for the PHANGS-HST-H$\alpha$ Treasury program (Cycle~30, PID 17126; PI: R. Chandar) were conducted starting in December 2022, and allocated Nineteen orbits. 
We note that NGC~2835-N was observed just prior to publication, so will be included in a future image release, and NGC\,4535 was ultimately observed as part of a related HST program (PID 17457, PI: Belfiore).
Imaging is conducted with the WFC3 camera in either the F658N or F657N filter (more below). PHANGS–H$\alpha$ footprint coverage is designed to maximize overlap with the five broad-band filters observed as part of the PHANGS-HST Treasury Survey (Cycle 26, PID 15654; PI: J. Lee).  We restrict the allowed orientation to match that of the Cycle 26  PHANGS-HST broad-band observations for cases where such restrictions do not make the visit too challenging to schedule.  There are either one or two pointings per galaxy.

\begin{figure*}[!htbp]
    \centering
        \includegraphics[width=\textwidth]{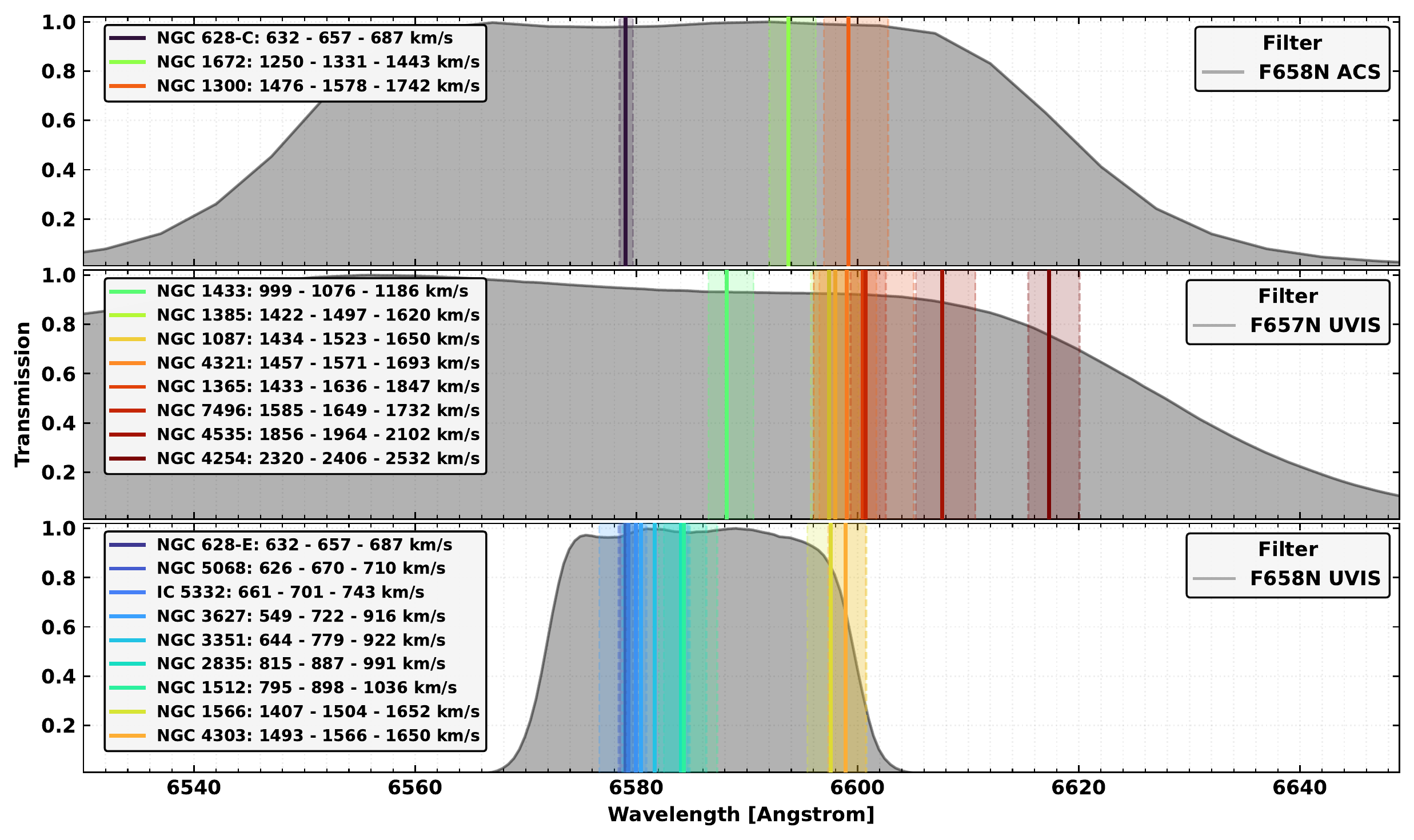}
    \caption{Filter transmission curves for the F657N (shown in middle panel) and F658N (shown in bottom panel) HST-H$\alpha$ filters on WFC3/UVIS used in this survey. The solid vertical lines show the wavelength of the H$\alpha$ line for the systemic velocity of each galaxy in our sample, while the shaded regions show the full velocity range.  Observations with the narrow F658N filter are preferred, but H$\alpha$ emission for galaxies with velocities higher than $\approx1500\,$km/s is redshifted outside of this passband, and we use F657N instead (see Fig.\,\ref{fig_Ha_ratio1}; see also Fig.\,\ref{fig_appendix1} for a comparison of HST to MUSE \Ha\ flux as a function of velocity for two galaxies that have a velocity range close to the edge of the filter transmission curve). 
    Archival observations for three galaxies (NGC~628-C, NGC1300, and NGC~1672) use the F658N filter from the ACS/WFC camera (shown in top panel).}
    \label{fig:filter}
\end{figure*}

There are two WFC3/UVIS narrow-band filters which cover the H$\alpha$ line at the distances of our sample galaxies: F658N and F657N.  The F658N filter is preferred because it is significantly narrower, with a width of 27.5~\AA~ or $\sim1300$~km~s$^{-1}$ (and a pivot wavelength of 6585.6~\AA), compared with the F657N filter which has a width of 121~\AA~ or $\sim5500$~km~s$^{-1}$ (and a pivot wavelength of 6566.6~\AA). Note that even the narrow F658N filter includes the [NII]$\lambda$6583 emission line.  We show the transmission curves for the two filters in Figure~\ref{fig:filter}.

Both filters have a peak throughput of 25\%. We use the narrow F658N filter to observe galaxies with systemic velocities less than $\approx1500$~km/s and the F657N filter to observe those with higher velocities  (note that the F657N filter was used to image NGC~1433 even though it has a velocity $\lea 700$~km~s$^{-1}$).

With the WFC3/UVIS camera, we observe a single target in each visit, and maximize the integration time in each exposure.  
A post flash of 17 or 18 e$^-$ is applied for each exposure to increase the background counts, 
which is recommended to mitigate issues due to charge transfer efficiency (CTE) losses. The dither sequence is optimized to cover the WFC3 chip gap. We obtain two exposures at each of two pointings using the `WFC3-UVIS-GAP-LINE' pattern, which is designed to cover the gap between the two WFC3 CCDs. The exposure times for each galaxy are listed in Column~4 of Table~\ref{tab:galaxysample}.
\begin{table*}[!htbp]
\caption{Summary of HST Observations and MUSE parameters.}
\centering
\begin{tabular}{llllll}
\hline
\hline

\textbf{Galaxy} & Velocity & Filter & ExpTime & MUSE Res. \\ 
&  [km~s$^{-1}$] & Instrument & [sec]  & [arcsec] \\ 
(1) & (2) & (3) & (4) & (5) \\ 
\hline
IC~5332  & $701\pm1$	& F658N WFC3 & $4\times598$ & 0.87\\ 
NGC~628-C*   & $657\pm1$	& F658N ACS & $2\times711$	& 0.92\\ 
NGC~628-E    & $657\pm1$	& F658N WFC3 & $4\times587$	& 0.92\\ 
NGC~1087    & $1523\pm2$	& F657N WFC3 & $4\times584$	 & 0.92\\ 
NGC~1300*   & $1578\pm1$ &  F658N ACS & $4\times680$ & 0.89 \\
NGC~1365     & $1636\pm1$	& F657N WFC3 &  $4\times598$ & 1.15\\ 
NGC~1385    & $1497\pm4$	& F657N WFC3 & $4\times 550$ & 0.77	\\ 
NGC~1433* & $1076\pm1$ & F657N WFC3 & $3\times515$ & 0.91 \\
NGC~1512    &	$898\pm3$ &	F658N WFC3 & $4\times602$ & 1.25   \\
NGC~1566    & $1504\pm2$	&	F658N WFC3 & $4\times632$  & 0.80\\ 
NGC~1672*    &	$1331\pm3$ & F658N ACS & $1\times593 , 3\times617$  & 0.96\\ 
NGC~2835-S    	& $887\pm1$ &	F658N WFC3 & $4\times588$  & 1.15  \\ 
NGC~2835-N    	& $887\pm1$ &	F658N WFC3 & $4\times556$  & 1.15  \\ 
NGC~3351     & $779\pm1$	&	F658N WFC3 & $4\times586$  & 1.05  \\ 
NGC~3627     &$722\pm2$	& F658N WFC3 & $4\times550$	 & 1.05   \\ 
NGC~4254-E     &$2406\pm1$	&	F657N WFC3 & $4\times586$ & 0.89  \\ 
NGC~4254-W    &$2406\pm1$	&	F657N WFC3 & $4\times586$  & 0.89   \\ 
NGC~4303   &$1566\pm2$	&  F658N WFC3 & $4\times584$	&0.78 \\ 
NGC~4321    &	$1571\pm1$& F657N WFC3 & $4\times587$	&1.16 \\ 
NGC~4535     &$1964\pm1$	&	F657N WFC3 & $4\times550$ & 0.56   \\ 
NGC~5068-N    &$670\pm1$	&	F658N WFC3 & $4\times588$ & 1.04  \\ 
NGC~5068-S     &	$670\pm1$&	F658N WFC3 & $4\times588$ & 1.04 \\ 
NGC~7496    & $1649\pm5$ &	F657N WFC3 & $4\times608$ &0.89 \\ 
\hline
\hline
\end{tabular}\\
\tablecomments{Archival observations were taken by the following programs: GO-10402 (PI: Chandar) for NGC~628C, GO-10342 (PI: Noll) for NGC~1300, GO-13773 (PI: Chandar) for NGC~1433, and GO-10354 (PI: Jenkins) for NGC1672. Column~5 lists the FWHM of the Gaussian PSF of the homogenised (copt) mosaic from \citet{Emsellem2022}}
\label{tab:galaxysample}
\end{table*}
We show footprints of the new PHANGS-HST-H$\alpha$ coverage (red) and that of VLT/MUSE IFU (white) in the Appendix.

Standard procedures are used to reduce and drizzle the F658N and F657N images. The exposures are processed through the Pyraf/STSDAS CALWFC3 or CALACS software, which performs initial data quality flagging, bias subtraction, gain correction, bias stripe removal, correction for CTE losses, dark current subtraction, flat-fielding and photometric calibration, and outputs ``FLC'' FITS files for each exposure. These files are then aligned and registered with North up and East to the left, sky subtracted, and drizzled (using the DRIZZLEPAC software package) to create a single F658N or F657N image matched to the pixel grid of the PHANGS-HST V band images (Lee et al. 2022), with a pixel scale of $0.04\arcsec$.  The PHANGS-HST V-band images used astrometry calibrated using GAIA DR2 sources (Gaia Collaboration et al. 2018; Lindegren et al. 2018). The final FITS files are in units of e$^-$~s$^{-1}$ and important use-cases are discussed in \S\ref{sec:dataproducts}.

We used archival narrow-band images for NGC~628C, NGC~1300, NGC~1433, and NGC~1672. These were taken by the following programs: GO-10402 (PI: Chandar) for NGC~628C, GO-10342 (PI: Noll) for NGC~1300, GO-13773 (PI: Chandar) for NGC~1433, and GO-10354 (PI: Jenkins) for NGC1672.  The specific filter and exposure times are compiled in Table~\ref{tab:galaxysample} for these data as well. 


As noted in the previous Section, two other HST programs (besides PID 17126) continue to obtain broadband and H$\alpha$ imaging of PHANGS galaxies that were not part of the PHANGS-JWST Cycle 1 Treasury sample.  Specifically, PID~17457 (PI: F. Belfiore) is completing H$\alpha$ imaging for the rest of the original 38 PHANGS-HST galaxies.  Additionally, PID~17502 (PI: D. Thilker) a Cycle 31 Large Treasury is obtaining both NUV-U-B-V-I and H$\alpha$ for PHANGS galaxies observed by JWST in Cycle 2.  When completed these programs will expand science analysis such as described herein to 74 galaxies, each with HST+JWST+ALMA datasets.  The PHANGS collaboration is working to obtain MUSE IFU spectroscopy for as many of these as possible.

\subsection{MUSE Observations}\label{sec:MUSEdata}

The PHANGS-MUSE survey has carried out observations with the MUSE integral field spectrograph at the ESO VLT to map the 19 nearby spiral galaxies that form our sample \citep{Emsellem2022}. The footprints of the PHANGS-MUSE survey were designed to maximize overlap with the PHANGS-ALMA CO(2-1) maps, and are shown in the Appendix (Figure~\ref{fig:footprints}). 
The spatial resolution of the MUSE observations for each galaxy are compiled in Table~\ref{tab:galaxysample}. PHANGS-MUSE has produced $\approx15 \times10^6$ spectra which have a wavelength coverage of  4800-9300~$\AA$ with a spectral resolution going from $\sim80$~km~s$^{-1}$ at the blue end to $\sim35$~km~s$^{-1}$ at the red end, and 
targeting key optical lines including H$\beta$, [OIII]$\lambda5006$, [OI]$\lambda6300$ [SII]$\lambda6716$,$\lambda6731$, H$\alpha$, and [NII]$\lambda6583$.


The data reduction process is described in detail in Emsellem et al. (2022).  The data cubes are publicly available from the ESO archive\footnote{https://www.eso.org/sci/publications/announcements/sciann17425.html}.  In this work, we use the PHANGS-MUSE spectra to help anchor the flux calibration of our HST-H$\alpha$ images, and to correct for [NII] contamination, as described in Section~\ref{sec_fluxan}.


\section{Processing of PHANGS-HST-H$\alpha$ Images}
\label{sec_fluxan}

\begin{figure*}[!htbp]
    \centering
        \includegraphics[width=\textwidth]{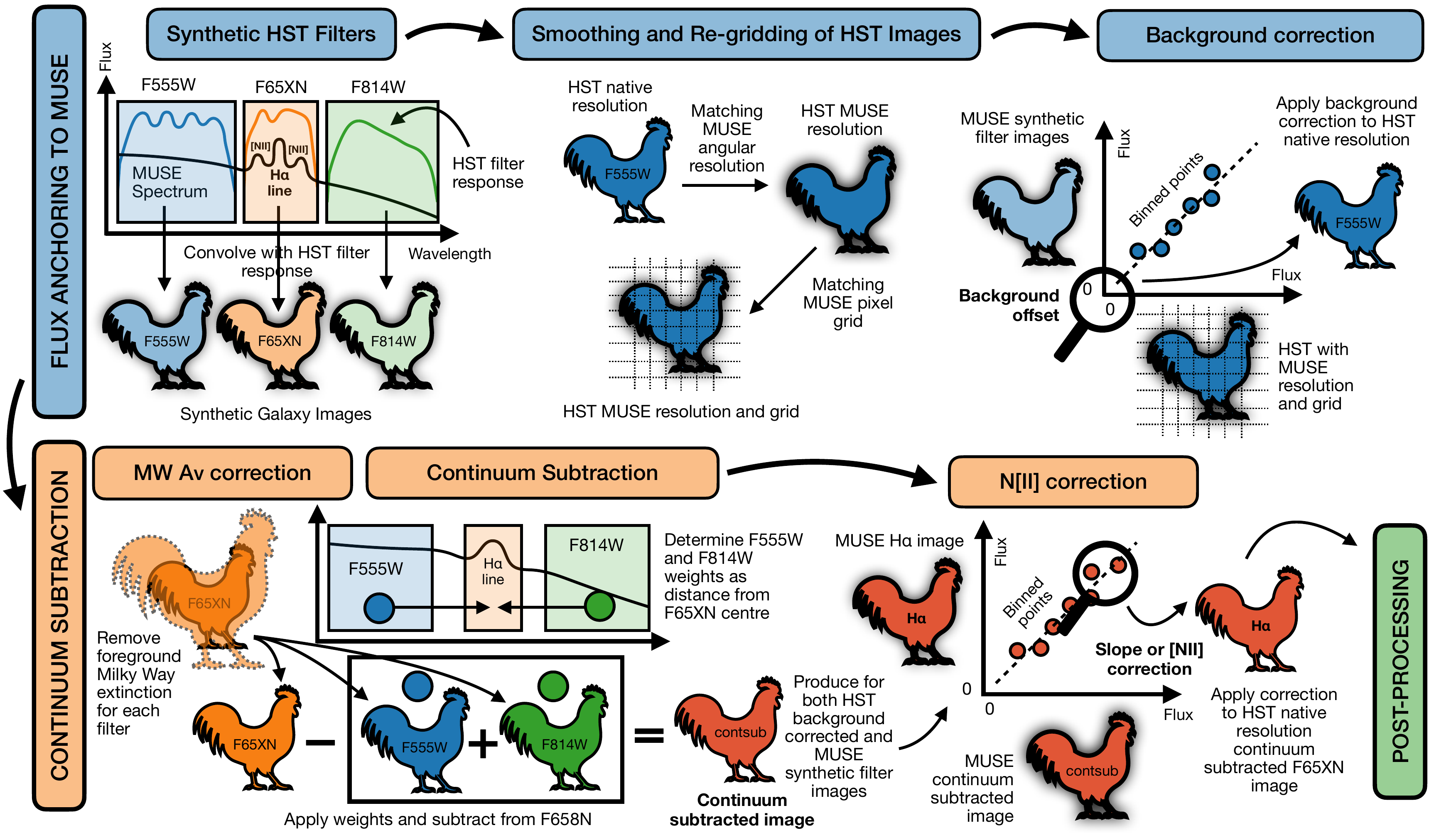}
    \caption{{\bf Overview of the \texttt{PyHSTHAContSub} (\citealp{ashley_barnes_2025_14610187}) pipeline background and continuum subtraction steps used to produce the HST H$\alpha$ images (see \S\,\ref{sec_fluxan}).} In our workflow, we first generate synthetic HST filter images from the MUSE data (\S\,\ref{subsec_fluxan1}). We then smooth and regrid the HST images (while preserving flux) to the resolution and pixel grid of the synthetic MUSE images. We compare these images and determine a background correction, which we apply to the full resolution HST images. For the continuum subtraction, we first subtract the Milky Way foreground extinction from each filter (both for the MUSE and HST data). We then determine the weights of the two broadband images, and subtract a combination of these two from the narrow-band image to produce the continuum-subtracted narrow-band images. We do this for both the HST and MUSE observations. We then compare the MUSE H$\alpha$ image with the continuum-subtracted MUSE image to determine the contribution of [NII] line emission in the narrow-band filter, which we then apply to the full-resolution HST images. These final images are then passed to the post-processing steps to correct for oversubtraction, remaining cosmic ray and foreground star residuals. Here the chickens represent the individual galaxy images.}
    \label{fig_flowchart}
\end{figure*}
 In this section, we describe our \texttt{PyHSTHAContSub} (\citealp{ashley_barnes_2025_14610187}) pipeline for producing flux-calibrated, continuum-subtracted high-resolution HST-H$\alpha$ maps
 (see flowchart in Fig.\,\ref{fig_flowchart}). 
There are a number of challenges in determining and correcting for the background level (which is not well constrained for the HST images) and correcting for contamination from [NII] line emission that is included in the passband.
For these purposes, we leverage our MUSE/IFU spectroscopy \citep{Emsellem2022}, which provides direct constraints on the [NII] contamination (convolved with the HST filter transmission) and has been calibrated to correctly match the background level. Post-processing steps correct for remaining artifacts, including cosmic ray residuals, foreground stars, and strongly negative regions where the continuum has been over-subtracted.

\subsection{Synthetic MUSE Images}
\label{subsec_fluxan1}

\begin{figure*}[!htbp]
    \centering
    \includegraphics[width=\textwidth]{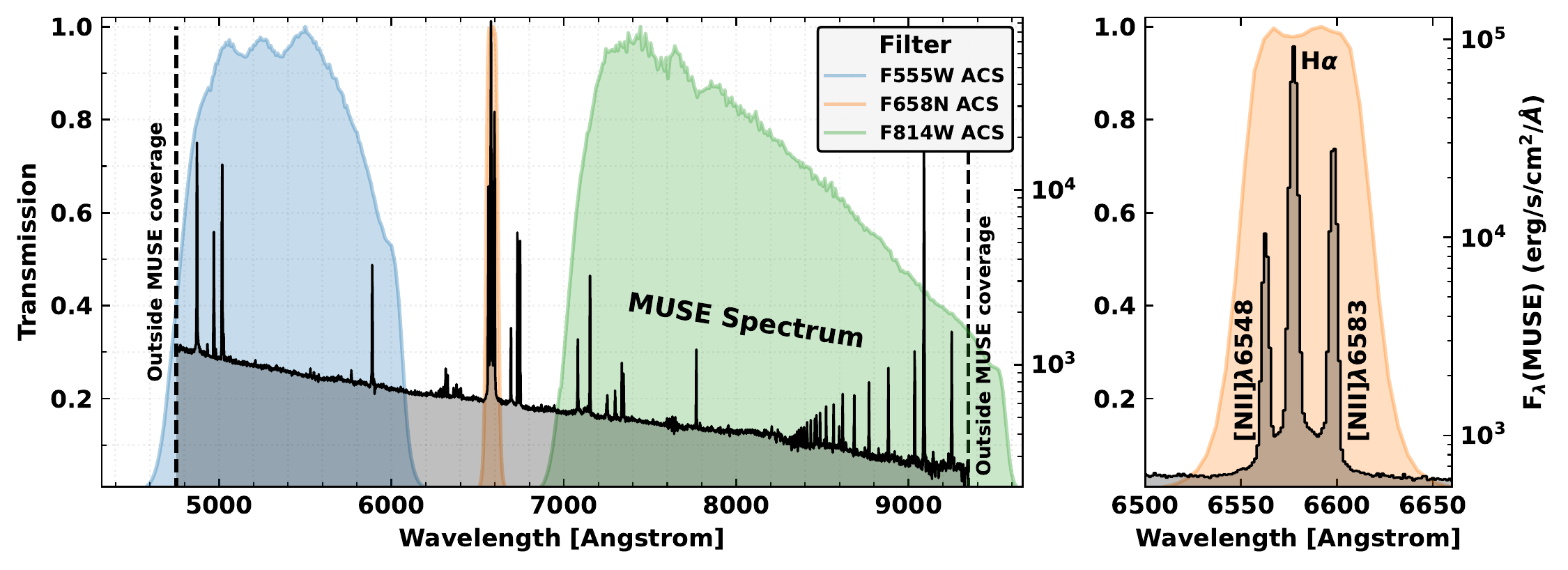} \\ \vspace{-6mm}
    \includegraphics[width=\textwidth]{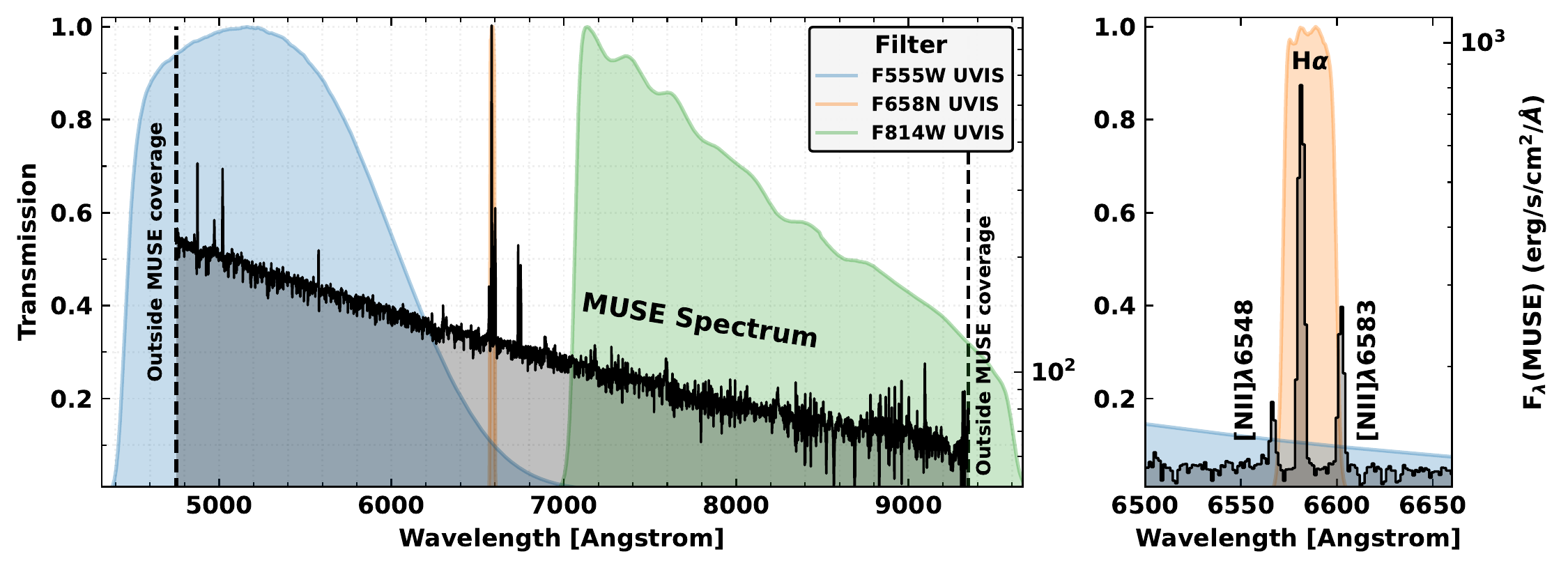}
    \caption{\textbf{Example MUSE spectra compared with HST filter transmission curves (ACS/WFC and WFC3/UVIS).} 
    Example spectra from NGC\,628 (top panel) and NGC\,2835 (bottom panel) are shown as solid black lines, overlaid with the F555W (blue), F658N (orange), and F814W (green) filter transmission curves for ACS/WFC (top panel) and WFC3/UVIS  (bottom panel). The left panel displays the full wavelength range of the MUSE observations, with dashed vertical lines marking the coverage limits. In the right panel, we focus on the \Ha\ line within the F658N filter bandpass, which includes contributions from the [NII]$\lambda6548$ and [NII]$\lambda6583$ emission lines.}
    \label{fig_filters_spec}
\end{figure*}

We generate synthetic images from the MUSE data cubes integrating over the HST F657N, F658N, F555W, and F814W filters associated with the appropriate camera$+$ filter (WFC3/UVIS or ACS/WFC) for each galaxy (see Appendix\,\ref{appendix_555filters} for comparison of alternative filter selection).

The wavelength coverage of the MUSE spectra does not fully cover the range of the F555W and F814W transmission curves (see  Fig.\,\ref{fig_filters_spec}). 
Consequently, a small fraction of the emission remains unaccounted for in the synthetic images. However, the continuum subtraction is performed in units of flux density (rather than flux, see below), so the small missing bandwidth does not significantly affect 
the final HST-H$\alpha$ maps.

\subsection{Convert Images to Flux Density}
\label{subsec_fluxan2}

The HST and synthetic MUSE images are all converted to the same units of flux density (\fluxdensityunitpix). To convert the HST images from units of electrons/s to flux density, we employ the \texttt{photflam} parameter. The \texttt{photflam} value is for a source with a constant flux per unit wavelength in units of \fluxdensityunit\ that produces a count rate of one count per second, and is taken from the header of the HST images. Additionally, we use the \textsc{synphot} package\footnote{\url{https://synphot.readthedocs.io/en/latest/}} to calculate \texttt{photplam}, or the pivot wavelength (`pivot' parameter), and the \texttt{photbw}, or width of the filter (`rectwidth' parameter). These parameters for all filters are summarized in Tab.\,\ref{tab_photparams}.
\begin{table}

    \centering
    \caption{Parameters taken from {\sc synphot.SpectralElement}. These are used to convert the filter images from units of counts\,s$^{-1}$ to units of flux density (\S\,\ref{subsec_fluxan2}) and the H$\alpha$ images into flux (\S\,\ref{subsec_contsub1}).}
    \label{tab_photparams}

\begin{tabular}{ccccc}
\hline\hline
Instrument & Filter & photflam & photplam & photbw \\
 &  & $\mathrm{\,cm^{2}\,\mathring{A}\,s\,erg^{-1}}$ & $\mathrm{\mathring{A}}$ & $\mathrm{\mathring{A}}$ \\ 
&  & $\times \mathrm{10^{-19}}$ & & \\ 
\hline 
ACS/WFC & F555W & 1.988 & 5361.0 & 1124.6 \\
ACS/WFC & F658N & 20.063 & 6584.0 & 74.9 \\
ACS/WFC & F814W & 0.712 & 8045.5 & 1741.5 \\
WFC3/UVIS & F555W & 1.827 & 5308.4 & 1565.4 \\
WFC3/UVIS & F657N & 21.814 & 6566.6 & 121.0 \\
WFC3/UVIS & F658N & 97.583 & 6585.6 & 27.5 \\
WFC3/UVIS & F814W & 1.499 & 8039.0 & 1565.2 \\
\hline \hline
\end{tabular}

\end{table}

\subsection{Match HST Images to MUSE Resolution}
\label{subsec_fluxan3}

\begin{figure*}[!ht]
    \centering
        \includegraphics[width=\textwidth]{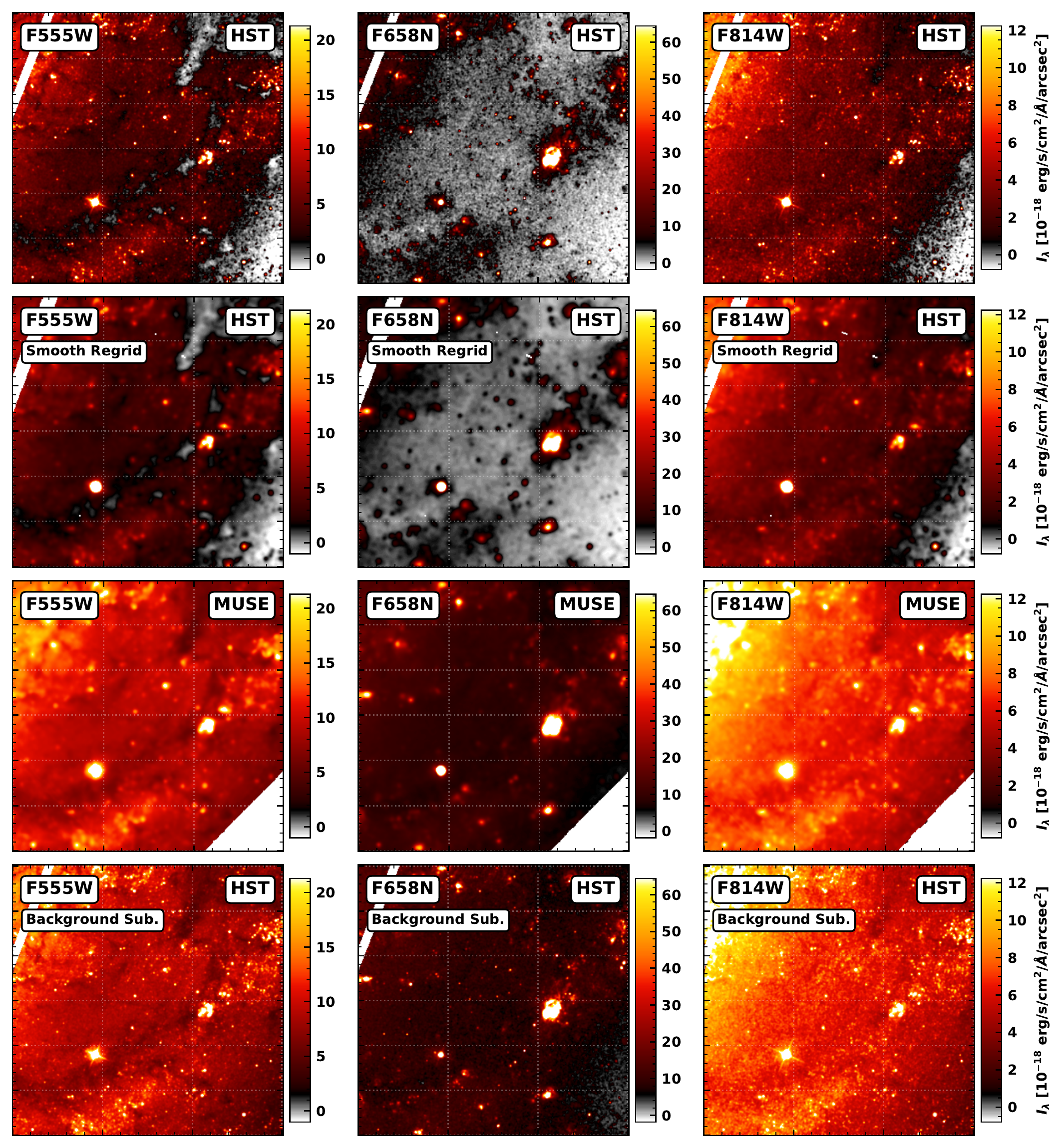}
    \caption{\textbf{HST and MUSE Filters for NGC\,628.} 
    \textit{(First row, left to right)}: HST F555W, F658N, and F814W filters at their native resolution ($\sim$0.1\,arcsec) and pixel grid. 
    \textit{(Second row, left to right)}: HST filters smoothed to the resolution and grid size matching the MUSE observations (see next row of panels). 
    \textit{(Third row, left to right)}: Synthetic images produced by applying the HST transmission curves to the lower resolution MUSE spectroscopic observations.  These are used to improve the flux-calibration of the HST images. 
    \textit{(Fourth row, left to right)}: HST filter maps (again at native resolution and pixel grid) with zero-point offsets applied; offsets determined by comparing with the synthetic MUSE images (see Fig\,\ref{fig_spectrum}). 
    }
    \label{fig_maps}
\end{figure*}

Next, we smooth and regrid the HST images to match the resolution and pixel grid of the synthetic MUSE images (see Fig.\,\ref{fig_maps}).
We first smooth all HST images using a Gaussian function with a size of,
\begin{equation}
\theta_\mathrm{kernel} = (\theta_\mathrm{MUSE}^2 - \theta_\mathrm{HST}^2)^{0.5}, 
\end{equation}
where the resolution of the MUSE observations ($\theta_\mathrm{MUSE}$) for each galaxy are compiled in Table\,1 (see \citealp{Emsellem2022}), and the HST resolution ($\theta_\mathrm{HST}$) is assumed to be 0.1\arcsec. In practice, the point spread function of the HST observations is not necessarily a Gaussian \citep{Krist11}, and will fluctuate by $\approx0.02\arcsec$ around this value depending on the filter and pixel coordinates.  However, since the HST images have approximately an order of magnitude higher spatial resolution than the MUSE observations, and are, hence, smoothed significantly during this step of the procedure, small variations in the PSF/shape of the beam of the HST images have essentially no impact on the resulting MUSE-resolution-matched HST-H$\alpha$ maps.

We convolve the smoothed HST images to the same pixel-grid as the synthetic MUSE images (pixel size of 0.2\arcsec) using a bilinear function (making use of the {\sc astropy.convolution.convolve$\_$fft} function; \citealt{AstropyCollaboration2018}). 

\subsection{Background Level from MUSE Observations}
\label{subsec_fluxan4}

We compare the flux density values in each pixel of the matched-resolution HST and synthetic MUSE images. The best linear fit between these will be used to anchor fluxes in the HST images to the more accurate MUSE values. To determine this relation, we first mask pixels that are: (1) contaminated by bright foreground stars (identified from MUSE; \citealp{Groves23}), and (2) limit the data to a range around zero. The latter was done since in this step we are primarily interested in the calibration of any zero-point offset between the two images and therefore want to limit the dependence of our fitting on the bright emission regions. We limit the data between 0.01\% and 99\% percentiles in the HST and synthetic MUSE images. In addition, we bin the flux data into 25 bins, which again limits the impact of outlier pixels on the fit. A similar method was recently employed for the calibration of the PHANGS-JWST images \citep{Leroy2023,Williams24}. Lastly, a linear least square fitting ($y = ax + b$, where $y$ represents the HST image data points and $x$ represents the MUSE image data points) was performed. 

\begin{figure*}
    \centering
        \includegraphics[width=\textwidth]{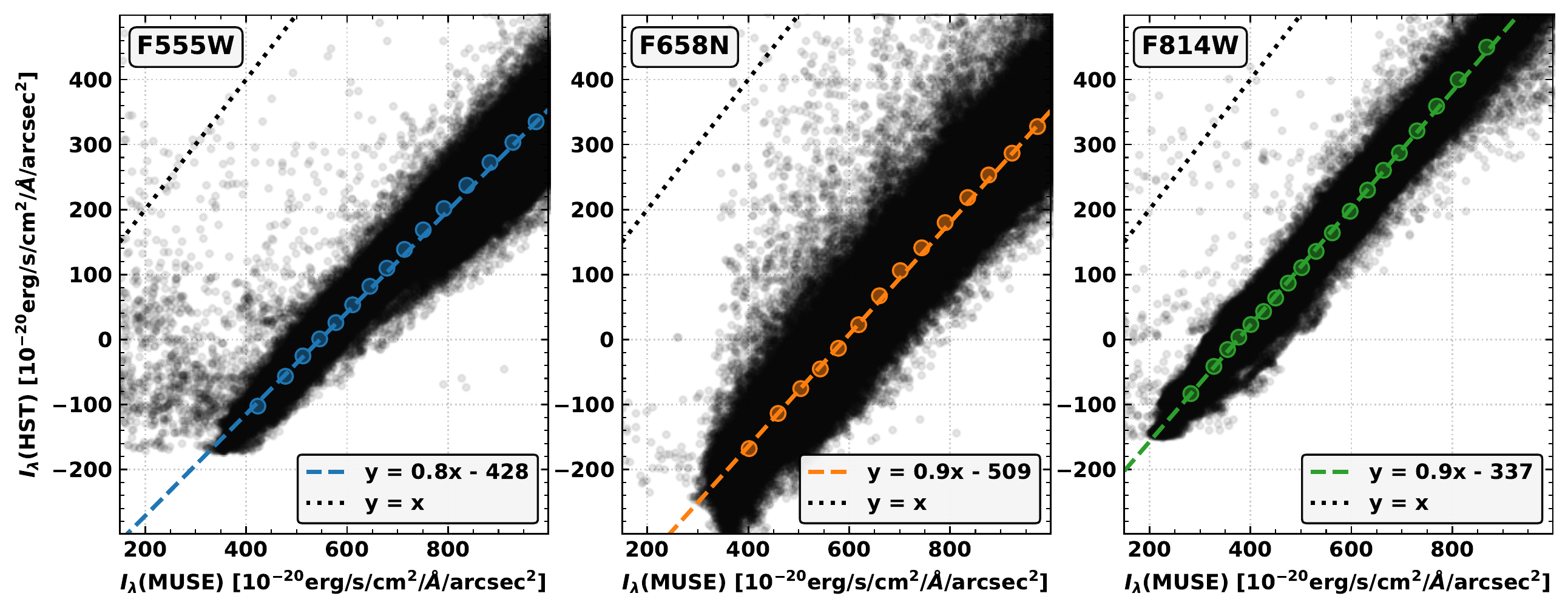}
    \caption{\textbf{
    Fits to determine the background level in HST images by calibrating to MUSE observations (\S\,\ref{subsec_fluxan4}).} 
    Intensities from matched-resolution pixels in the HST versus synthetic MUSE images measured in the F555W, F658N, and F814W filters for NGC~628-C.
The colored circles represent 25 binned data points (spaced by data density), and the dashed line in each panel shows the best linear fit to these points ($y = ax + b$), where $y$ represents measurements from the degraded HST pixels and $x$ represents the MUSE pixels (see Tab.\,\ref{tab_bgcorr_niicorr}). The dotted line in each panel shows the 1-to-1 line.  The best fit offsets are subtracted from the full-resolution HST images to correct for the sky background level.}
    \label{fig_spectrum}
\end{figure*}

The results of this analysis can be seen in Fig.~\ref{fig_spectrum}, which shows the matched-resolution HST (y-axis) and synthetic MUSE (x-axis) flux densities for the F555W, F658N, and F814W filters towards NGC 628-C, as well as the binning and fitting routines.
We apply the background correction from these solutions to the full-resolution HST images as $y_\mathrm{bgsub} = y - b$ (i.e. we do not correct for the slope of the fit, but only the background offset).\footnote{In the Appendix we show that fixing the slope to unity has only a minor affect on the fitted background level
(see Tab.\,\ref{tab_bgcorr_niicorr_fixed} in the appendix).}
The background correction offsets for each galaxy and each filter are given in Tab.\,\ref{tab_bgcorr_niicorr}.
\begin{table}[!htbp]
\centering

    \caption{Results of fitting for the background correction in each of the HST filters (\S\,\ref{subsec_fluxan4}), and the correction factor for the [N\,II] $\lambda$ 6548, 83\AA\ line contamination in the H\,$\alpha$ images (\S\,\ref{subsec_contsub2}).}
    \label{tab_bgcorr_niicorr}
    
\begin{tabular}{ccccc}
\hline \hline
Galaxy & \multicolumn{3}{c}{Background Corrections} & [NII] Correction \\
 & \multicolumn{3}{c}{[erg\,s$^{-1}$\,cm$^{-2}$\,\AA$^{-1}$\,arcsec$^{-2}$]} & \\

 & F555W & F65XN & F814W & Factor \\
\hline
    IC~5332 & -290.78 & 1801.13 & -190.29 & 1.11 \\
    NGC~628-C & -428.03 & -509.67 & -337.13 & 1.35 \\
    NGC~628-E & -123.81 & 1637.38 & -196.70 & 1.19 \\
    NGC~1087 & -217.99 & 886.88 & -147.06 & 1.18 \\
    NGC~1300 & -84.71 & -92.37 & -51.87 & 1.26 \\
    NGC~1365 & -777.32 & 520.94 & -687.02 & 1.25 \\
    NGC~1385 & -242.34 & 890.50 & -164.83 & 1.21 \\
    NGC~1433 & -617.62 & -620.56 & -534.81 & 1.33 \\
    NGC~1512 & -161.94 & 1414.89 & -140.05 & 1.03 \\
    NGC~1566 & -1030.71 & 1660.55 & -575.82 & 1.03 \\
    NGC~1672 & -194.30 & -348.71 & -163.62 & 1.38 \\
    NGC~2835-S & -465.89 & 751.65 & -342.54 & 1.02 \\
    NGC~3351 & -180.02 & 542.77 & -164.60 & 1.07 \\
    NGC~3627 & -764.28 & 10.71 & -672.63 & 1.16 \\
    NGC~4254 & -87.41 & 120.74 & -205.08 & 0.94 \\
    NGC~4303 & -1153.58 & -386.42 & -848.63 & 0.97 \\
    NGC~4321 & -804.18 & 668.72 & -667.56 & 1.23 \\
    NGC~4535 & -610.03 & 672.52 & -487.48 & 1.11 \\
    NGC~5068 & -247.92 & -55.93 & -157.27 & 1.10 \\
    NGC~7496 & -201.02 & 506.70 & -146.29 & 1.22 \\
\hline
\end{tabular}
\end{table}
This procedure yields HST-H$\alpha$ maps with a robust background level anchored to match the MUSE measurements.

\subsection{Continuum Subtraction}
\label{subsec_contsub1}

The purpose of this step is to subtract the stellar continuum present in the narrow-band HST filter images (see Fig.\,\ref{fig_filters_spec}). To achieve this, we utilize the HST F555W and F814W images, which are dominated by the stellar continuum and have been flux-anchored to the synthetic MUSE images, as described in the previous section. We note that our procedure would also work with other filters. For example, the F550M or F547M filter could replace the F555W filter, as demonstrated in the Appendix.

We begin by correcting the observations for foreground extinction due to the Milky Way. 
This is achieved using the extinction law of \citet{cardelli89} and \(E(B-V)\) values from \citet{schlafly11}.
Although this adjustment results in a relatively minor flux correction (on the order of 10\%), it is important for accurate continuum subtraction. 
Without it, systematic over- or under-subtraction could occur across filters due to the wavelength dependence of extinction. 
Additionally, the same foreground extinction correction has been applied to the MUSE DAP \Ha\ maps, which are used to account for the contribution of the [NII] line, as detailed in \citet{Emsellem2022}. 

To estimate the continuum flux in the narrow-band F658N or F657N filter, we calculate weighting factors, \(W\), based on the relative differences in the central wavelengths of the broad- and narrow-band filters.  
\begin{equation}
    W_{\mathrm{F814W}} = \frac{\lambda_{\mathrm{F65XN}} - \lambda_{\mathrm{F555W}}}{\lambda_{\mathrm{F814W}} - \lambda_{\mathrm{F555W}}},
\end{equation}

\begin{equation}
    W_{\mathrm{F555W}} = \frac{\lambda_{\mathrm{F814W}} - \lambda_{\mathrm{F65XN}}}{\lambda_{\mathrm{F814W}} - \lambda_{\mathrm{F555W}}},
\end{equation}
where \(\lambda_{\mathrm{F555W}}\), \(\lambda_{\mathrm{F814W}}\), and \(\lambda_{\mathrm{F65XN}}\) are the central `pivot' wavelengths of the F555W, F814W, and either F657N or F658N filters, determined from their respective transmission filter response curves (see \S\,\ref{subsec_fluxan2}). The continuum image is produced in logarithmic space as,

\begin{equation}
    \log_{10}(F_{\mathrm{cont}}) = W_{\mathrm{F555W}} \log_{10}(F_{\mathrm{F555W}}) + W_{\mathrm{F814W}} \log_{10}(F_{\mathrm{F814W}}),
\end{equation}
where \(F_{\mathrm{F555W}}\) and \(F_{\mathrm{F814W}}\) are the pixel fluxes in the F555W and F814W images (in units of erg/cm\textsuperscript{2}/s/\AA). The continuum-subtracted H$\alpha$ image is then produced in linear space as,

\begin{equation}
    F_{\mathrm{H\alpha}} = F_{\mathrm{F65XN}} - F_{\mathrm{cont}},
\end{equation}
where \(F_{\mathrm{F65XN}}\) represents the pixel fluxes in either the F657N or F658N image. As a final step, we convert the continuum-subtracted images from units of flux density to units of flux using the filter bandwidth (from the respective transmission filter response curve; see \S\,\ref{subsec_fluxan2}). 

\begin{figure*}[!htbp]
    \centering
        \includegraphics[width=\textwidth]{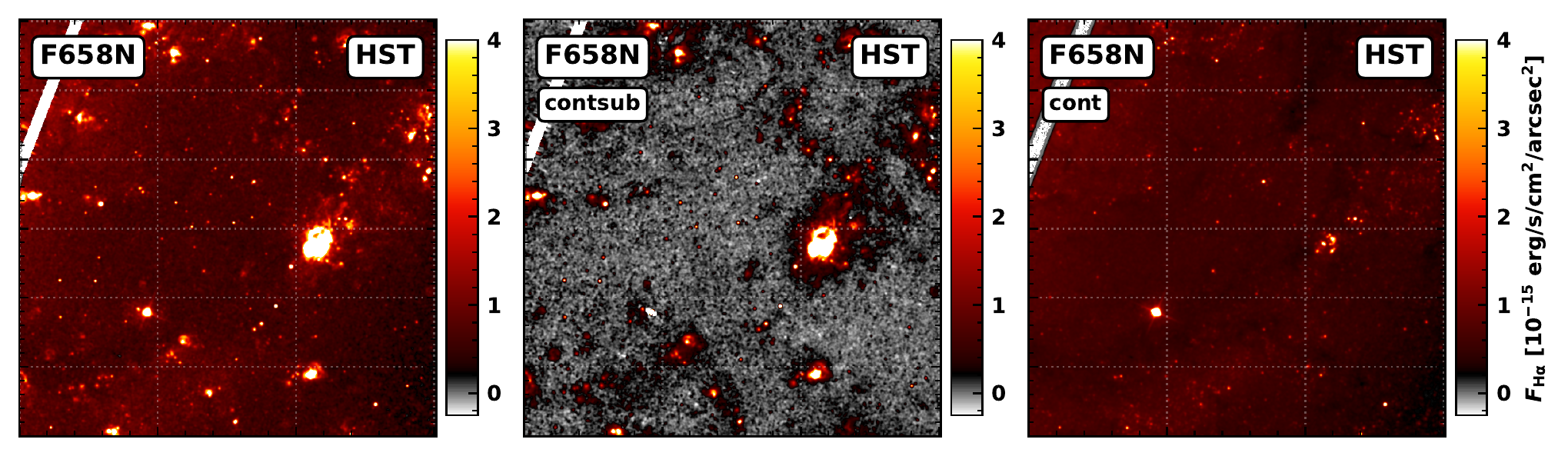} \\ \vspace{-2mm}
        \includegraphics[width=\textwidth]{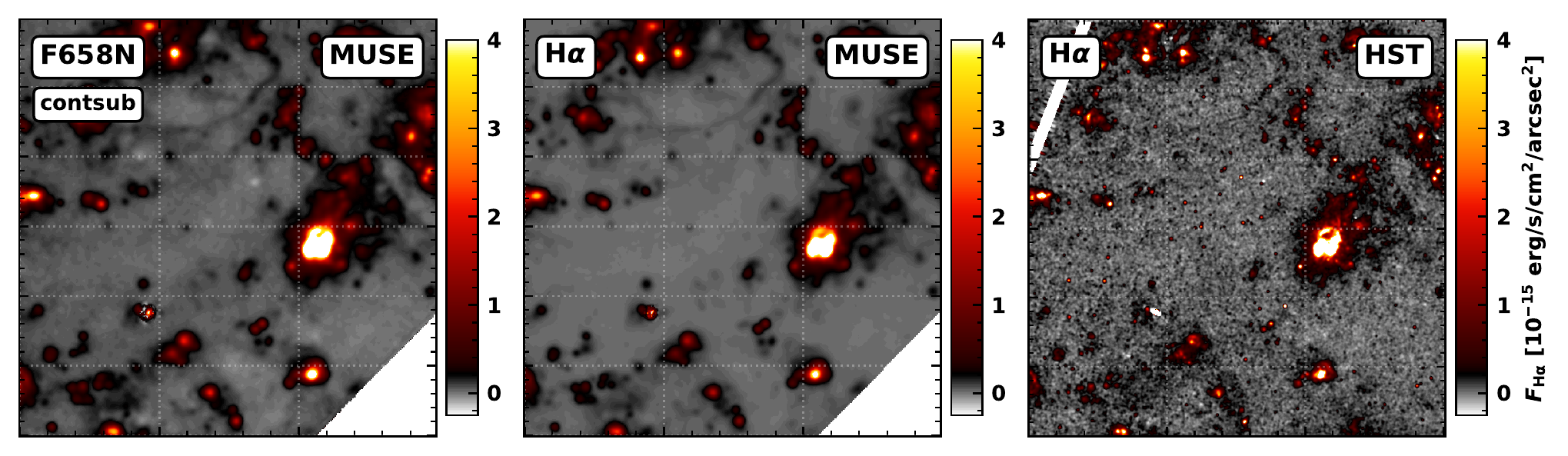}
    \caption{\textbf{Results of the extinction correction, continuum subtraction (\S\,\ref{subsec_contsub1}) and [NII] correction (\S\,\ref{subsec_contsub2}) of a portion of NGC~628-C (same as Fig.\,\ref{fig_maps}).}  (\textit{top panels from left to right}) Shown is the F658N filter image after background subtraction and extinction correction, and the continuum subtracted and continuum F658N filter images. \textit{(bottom panels from left to right)} The continuum subtracted F658N image produced from the synthetic HST filter MUSE images (i.e. including the contirbution from the [NII] lines convolved with the F658N filter transmission curve), the MUSE DAP \Ha\ emission only image (\citealp{Emsellem2022}), and the final [NII] corrected HST \Ha\ image.}
    \label{fig_mapscontsub}
\end{figure*}

The result of this continuum subtraction routine can be seen in Fig.\,\ref{fig_mapscontsub}.
Here, the bright nebulae are cleanly separated from the background stellar continuum across the whole disc of the galaxy (in this case NGC\,628). There are, however, low-level artifacts that are difficult to remove, e.g., due to diffuse stellar emission towards the center of the galaxy or compact stellar clusters that may have continuum spectra that are not well recovered by only using the (log) interpolation of the two broad-band observations. It is also worth noting that artifacts in either of the broadband or narrowband images become more prominent in the continuum-subtracted image, due to the relative intensity of the nebula emission to stellar emission.

\subsection{Correction for [N\,II] Emission}
\label{subsec_contsub2}

The final global correction applied to the HST-H$\alpha$ images accounts for the contribution from the [N\,II] $\lambda$6548\AA\ and [N\,II] $\lambda$6583\AA\ lines, which can fall within the narrow-band F658N and F657N filters depending on the HST filter and the velocity of the galaxies (see Fig.\,\ref{fig_filters_spec}). 
For this correction, we compare the HST-filtered MUSE images\footnote{Note that the HST-filtered MUSE images have already been corrected for foreground Galactic extinction.} to the DAP \Ha\ image to determine the correction factor. 
This factor is subsequently applied to the full-resolution continuum-subtracted HST observations. 
Using the MUSE data alone to derive this correction factor, rather than employing the smoothed HST observations, was motivated by the goal of conducting a consistent test using data from the same set of observations.

To do so, we conduct the continuum-subtraction analysis (including the MW extinction correction) outlined in \S\,\ref{subsec_contsub2} using the broadband- and narrow-band synthetic MUSE images for each galaxy. 
This produces MUSE continuum subtracted emission maps, with the H$\alpha$ and [NII] lines convolved with the filter transmission curves for each galaxy (see maps in lower panels of Fig.\,\ref{fig_mapscontsub}).\footnote{Note that these images were used, and not e.g. the ratio of the MUSE image H$\alpha$/([NII]$\lambda6548+$[NII]$\lambda6583$), as these synthetic MUSE images retain the variation in the filter transmission, due to the velocity field of the galaxies.}

\begin{figure}
    \centering
        \includegraphics[width=1\columnwidth]{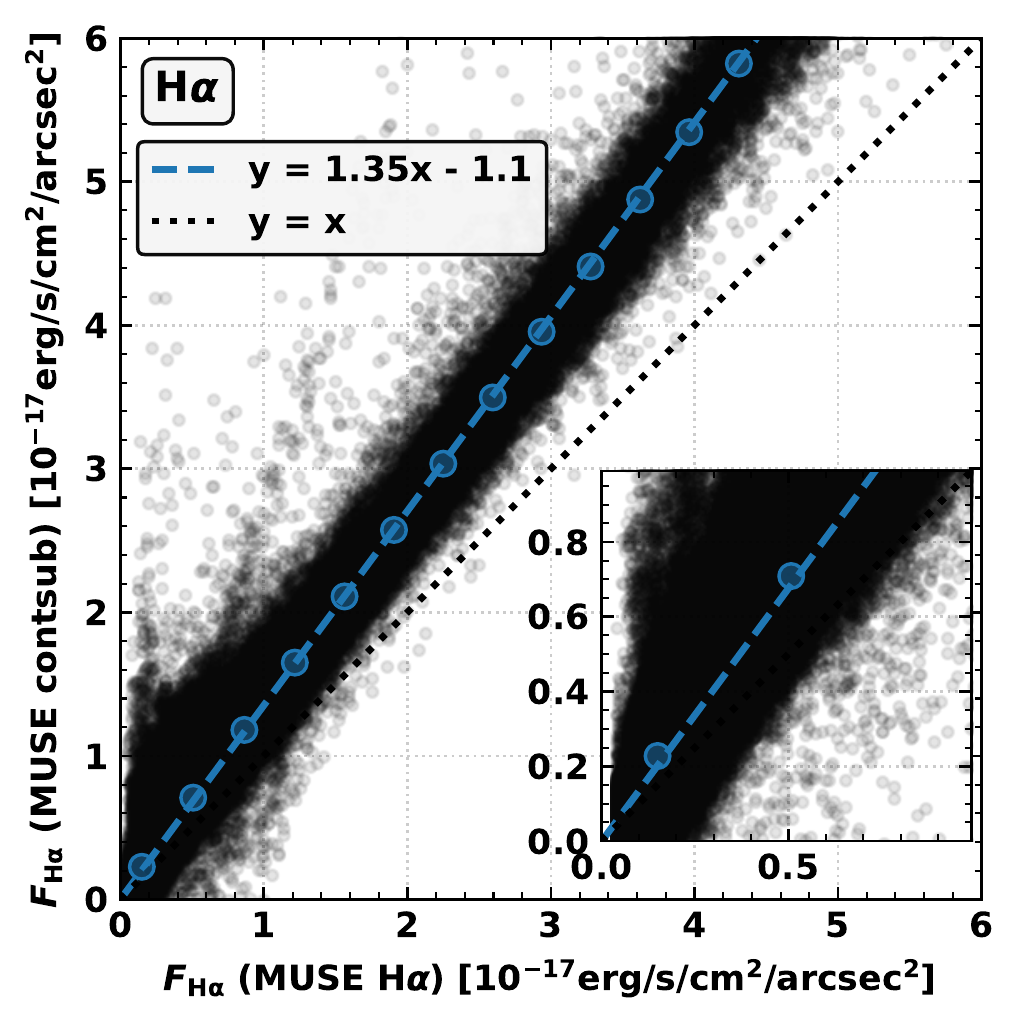}
    \caption{\textbf{Results of the [N\,II] $\lambda$6548\AA\ and [N\,II] $\lambda$6583\AA\ line correction routine (\S\,\ref{subsec_contsub2}).} H$\alpha$ fluxes from the continuum subtracted MUSE synthetic filter maps as a function of the H$\alpha$ flux densities measured from the MUSE spectroscopic data. The blue circles represent 20 binned data points (equally spaced), and the dashed blue line shows the best linear fit to these points ($y = ax + b$), where $y$ represents the HST image data points and $x$ represents the MUSE data points (see Table\,\ref{tab_bgcorr_niicorr}). The dotted line shows the 1-to-1 line.}\label{fig_scatteranchor}
\end{figure}

Next we compare the MUSE continuum-subtracted emission maps and the MUSE DAP \Ha\ emission line maps (see Fig.\,\ref{fig_scatteranchor}). 
We mask the outlying pixels with percentiles in both the MUSE continuum-subtracted and DAP \Ha\ maps (1 to 99 percent in this case), bin the remaining points into 20 equally spaced bins, and perform a least square fit to the binned data points assuming a linear polynomial ($y = ax + b$, where $y$ represents the MUSE continuum-subtracted emission data points and $x$ represents the MUSE DAP \Ha\ emission line data points; as shown in Fig.\,\ref{fig_scatteranchor}).
We then divide the full resolution HST fluxes by this factor ($a$), to obtain the HST \Ha-only emission line maps.\footnote{As we do not correct for the offset of the fit in this case (as opposed to the background subtraction, where we used a similar correction but only applied the offset; see \S\,\ref{subsec_fluxan4}), we also investigate the effect of fixing the offset to zero in this [N\,II] correction fitting procedure. However, doing so results in only a minor change in the correction (see Tab.\,\ref{tab_bgcorr_niicorr_fixed} in the appendix).}

It is worth noting that the relative contribution of these lines is somewhat challenging to constrain on a pixel-by-pixel basis across all galaxies. 
The exact contribution depends on the velocity, how the lines fall within the HST filter transmission curves, and the intrinsic nature and morphology of the specific source. 
For instance, assuming a global correction factor could introduce systematic errors in the flux with galactocentric radius, particularly in galaxies with strong metallicity gradients. 
In such cases, this approach might result in subtracting too little flux near the centre and too much flux near the edges. 
Indeed, we find that NGC\,4303 has a correction factor slightly below 1, which in principal should not be possible. 
This may be attributed to this issue, although this galaxy does not exhibit a particularly strong metallicity gradient \citep[see][]{Groves23}. 
This discrepancy could also arise from imperfect continuum subtraction, and/or absorption of the \Ha\ (which is corrected for in the MUSE DAP fits; see \citealp{Emsellem2022}).

Figure\,\ref{fig_scatteranchor} shows an example of this procedure for one galaxy, NGC\,628, where we find an [N\,II] correction factor of 1.35.
The final correction factors determined individually for each galaxy are compiled in Table\,\ref{tab_bgcorr_niicorr}. Again we note that fixing the intercept of the fits to zero makes only a minor change to the correction factors (see Tab.\,\ref{tab_bgcorr_niicorr_fixed} in the Appendix).

\begin{figure}[!t]
    \centering
        \includegraphics[width=\columnwidth]{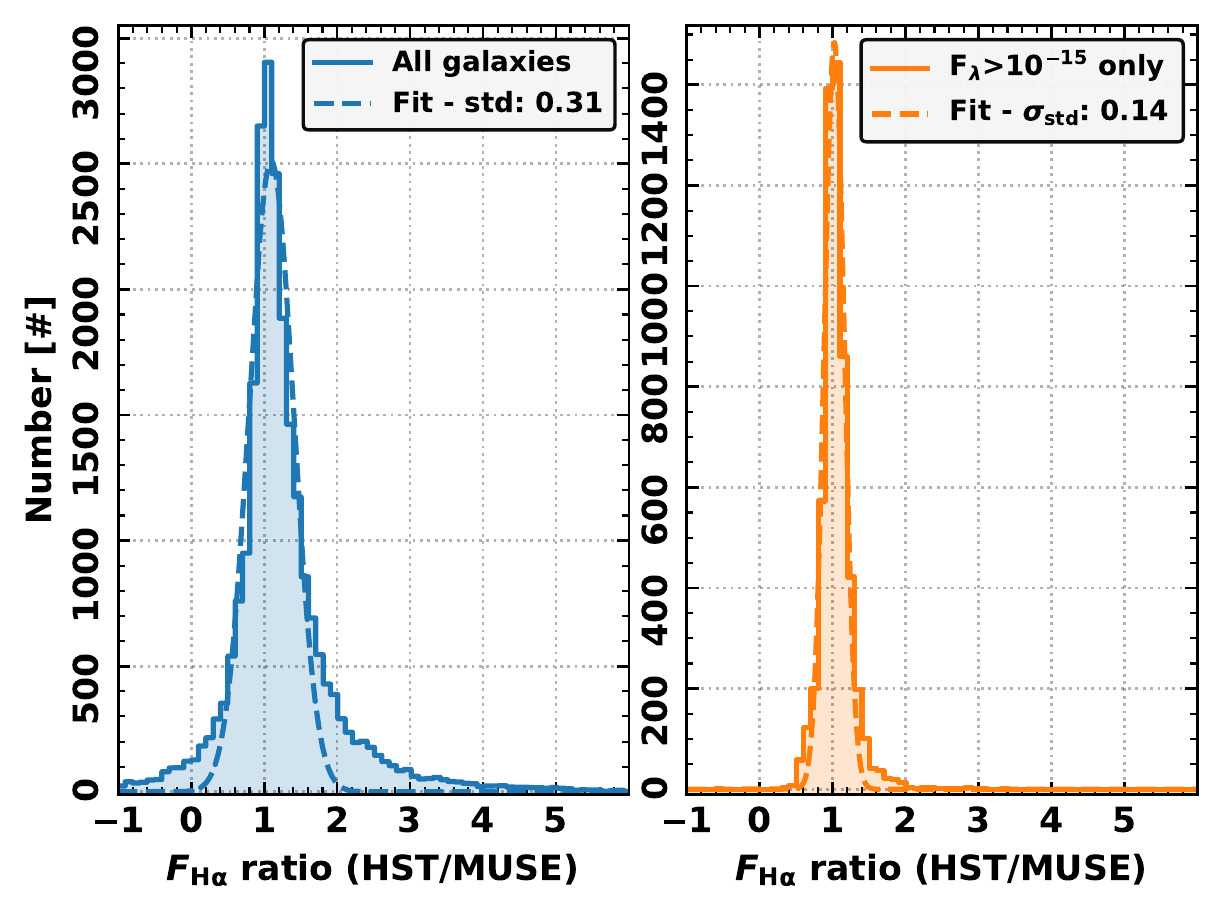}
    \caption{\textbf{Histograms of the final HST-to-MUSE H$\alpha$ flux ratios.} We show histograms of the HST to MUSE H$\alpha$ flux ratios for all regions (left panel) and for regions with MUSE H$\alpha$ fluxes higher than $10^{-15}$ erg/s/cm$^2$ (right). The dashed curves show the best fit Gaussian distribution in each panel, and the 1$\sigma$ standard deviation is written in the upper right.  For the bright H$\alpha$ regions, $\approx70$\% of the pixels in the HST-H$\alpha$ maps have fluxes within just $\approx$15\% of those measured from the MUSE spectra.}
    \label{fig_Ha_ratio1}
\end{figure}

We conduct a number of consistency checks of the resulting images. To do so, we sum the fluxes in the final HST-H$\alpha$ and MUSE H$\alpha$ maps within the masks from the MUSE nebulae catalogs. In Fig.\,\ref{fig_Ha_ratio1}, we plot the histogram distributions of this flux ratio for all sources within the catalog.
We see that the ratios scatter around 1, which highlights that the correction of the [N\,II] contribution in the HST data is reasonable. 
We note that we do not use the HST data to estimate this correction factor, but rather the synthetic MUSE observations, and despite this, the correlations are close to unity. We fit a Gaussian to the distribution for all galaxies and see that they have a standard deviation of around 30\%. 
Limiting the fit to the brightest regions above $10^{-15}$ erg/s/cm$^2$, we see this deviation decreases to around 15\%. 
We use these as the uncertainty confidence limits on our flux measurements across the HST H$\alpha$ images. 

\begin{figure*}[t]
    \centering
        \includegraphics[width=\textwidth]{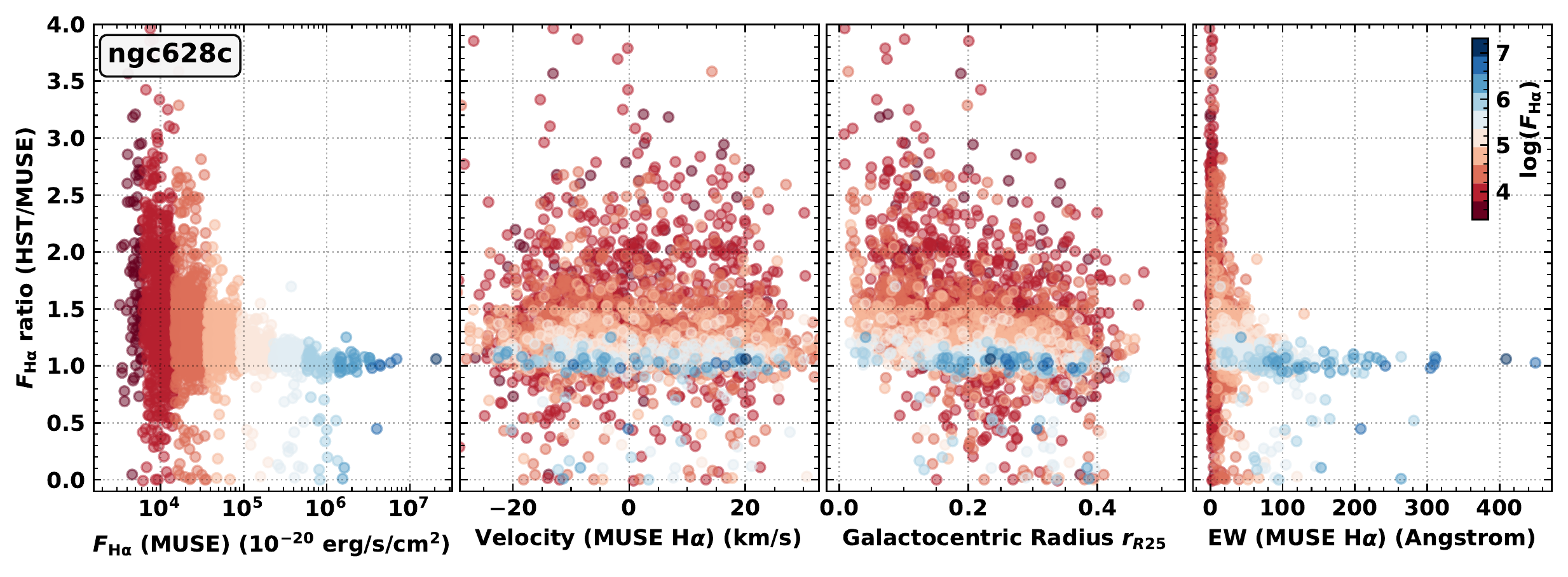}
    \caption{\textbf{The ratio of HST-to-MUSE H$\alpha$ fluxes in NGC~628-C are checked against other nebula properties.}
    We show (from left to right) the H$\alpha$ flux ratio versus the MUSE H$\alpha$ flux (left panel), MUSE H$\alpha$ velocity (second panel),  galactocentric radius (third panel), and the H$\alpha$ equivalent width measured directly from the MUSE spectra (right panel). There is no obvious trend in the flux ratio with H$\alpha$ velocity or galactocentric distance (middle two panels). 
    }
    \label{fig_Ha_ratio2}
\end{figure*}

In Fig.\,\ref{fig_Ha_ratio2}, we plot the ratio of these fluxes as a function of the MUSE H$\alpha$, MUSE H$\alpha$ velocity, galactocentric radius, and the MUSE H$\alpha$ equivalent width. 
Across the sample, we see evidence for mild systematic trends with galactocentric radius or velocity in a few cases, likely due to the emission shifting out of the filter (see Fig\,\ref{fig_appendix1}).

\subsection{Error Maps}
\label{subsec_errormaps}

To produce HST-H$\alpha$ error maps, we use the inverse variance ($1/V$) images for the individual filters (see \citealp{Lee2022}). These maps incorporate contributions from the detector and the Poisson noise towards target, modulated by the number of integrations drizzling into an output pixel and the effective exposure depth after masking, respectively. We convert the inverse variance image into a map of the uncertainties in units of counts per second (i.e. $\sigma = \sqrt{V}$), and then into flux density (\(\text{erg/cm}^2/\text{s/\AA}/\text{arcsec}^{2}\); see \S\,\ref{subsec_fluxan2}). To produce errors for the continuum subtracted narrow-band images, we propagate the individual pixel errors in all the bands as, 
\begin{equation}
    A = \left( W_{\mathrm{F555W}}\,\frac{\sigma_{{\mathrm{F555W}}}}{\mathrm{ln}(10)F_\mathrm{F555W}} \right)^2
\end{equation}
\begin{equation}
    B = \left( W_{\mathrm{F814W}}\, \frac{\sigma_{{\mathrm{F814W}}}}{\mathrm{ln}(10)F_\mathrm{F814W}} \right)^2
\end{equation}
\begin{equation}
    \mathrm{log_{10}(\sigma_{{\mathrm{cont}}})} = \sqrt{ A  + B}
\end{equation}
\begin{equation}
   \sigma_\mathrm{cont} = F_\mathrm{cont} \mathrm{ln}(10) \mathrm{log_{10}(\sigma_{{\mathrm{cont}}})}
\end{equation}
and then carry this forward to the continuum subtracted flux density as, 
\begin{equation}
    \sigma_{{\mathrm{H\alpha}}} = \sqrt{ \sigma_{{\mathrm{cont}}}^2 + \sigma_{{\mathrm{F65XN}}}^2}.
\end{equation}
This is converted into units of flux (see \S\,\ref{subsec_contsub1}) and corrected by the [NII] factor (see \S\,\ref{subsec_contsub2}). 
We note that these error maps do not account for uncertainties in the zero-point offset or in the [NII] correction factor, but are scaled accordingly by the correction factor such that they are representative of the errors in the final \Ha\ images. 
In Table\,\ref{tab_maprops} we provide some statistics of the noise across the HST-H$\alpha$ error maps. 

\begin{table*}[t]
\centering
\label{tab_maprops}
\caption{Noise statistics of the continuum-subtracted H$\alpha$ images and associated error maps for each galaxy. Compiled are the mean, and pixel values at 50 (median), 5, 16 (-$\sigma$), 84 (+$\sigma$), and 95 percentiles, and (median absolute) standard deviation for each map ($\sigma_\mathrm{STD}$).}
\begin{tabular}{l|ccccccc|ccccccc}
\hline \hline
Galaxy & \multicolumn{7}{c|}{Map Properties [10$^{-17}$\,erg/s/cm$^2$/arcsec$^{2}$]} & \multicolumn{7}{c}{Error Map Properties [10$^{-17}$\,erg/s/cm$^2$/arcsec$^{2}$]} \\
 & Mean & Median & 5\% & 16\% & 84\% & 95\% & $\sigma_\mathrm{STD}$ & Mean & Median & 5\% & 16\% & 84\% & 95\% & $\sigma_\mathrm{STD}$ \\
\hline
    IC~5332 & 6.5 & 4.0 & -92.4 & -54.1 & 64.6 & 108.0 & 59.3 & 77.9 & 73.4 & 66.5 & 69.1 & 80.4 & 93.2 & 4.9 \\
    NGC~0628-C & 15.4 & 8.9 & -31.4 & -15.4 & 36.8 & 64.1 & 25.6 & 43.3 & 41.4 & 36.7 & 38.5 & 46.2 & 55.1 & 3.4 \\
    NGC~0628-E & 5.9 & 3.3 & -99.3 & -58.3 & 67.6 & 114.7 & 62.8 & 82.1 & 77.2 & 69.9 & 72.7 & 85.0 & 101.4 & 5.2 \\
    NGC~1087 & 18.4 & 9.9 & -92.0 & -51.8 & 77.8 & 136.1 & 64.1 & 82.1 & 75.4 & 67.0 & 69.9 & 85.5 & 99.5 & 6.7 \\
    NGC~1300 & 4.0 & 2.4 & -22.5 & -12.5 & 18.2 & 30.7 & 15.2 & 27.6 & 26.4 & 23.6 & 24.7 & 29.0 & 34.0 & 2.0 \\
    NGC~1365-N & 29.9 & 13.9 & -79.6 & -42.4 & 76.4 & 135.1 & 58.6 & 75.0 & 68.8 & 60.9 & 63.6 & 79.6 & 95.7 & 6.5 \\
    NGC~1385 & 18.6 & 8.0 & -93.7 & -53.4 & 76.0 & 138.4 & 63.9 & 80.3 & 74.1 & 65.8 & 68.8 & 84.8 & 100.8 & 6.5 \\
    NGC~1433 & 14.0 & 3.9 & -92.4 & -53.8 & 64.7 & 110.5 & 59.1 & 82.1 & 75.7 & 66.6 & 70.0 & 86.3 & 102.8 & 7.0 \\
    NGC~1512 & 5.3 & 2.3 & -102.5 & -60.3 & 67.3 & 114.5 & 63.7 & 84.8 & 77.1 & 69.1 & 72.1 & 85.9 & 102.1 & 5.8 \\
    NGC~1566 & 26.7 & 11.3 & -91.1 & -50.5 & 80.4 & 145.1 & 64.6 & 81.7 & 75.0 & 67.0 & 69.9 & 84.2 & 99.5 & 6.1 \\
    NGC~1672 & 21.3 & 6.4 & -29.8 & -15.2 & 31.9 & 62.3 & 22.8 & 37.2 & 36.8 & 27.2 & 29.4 & 41.7 & 48.9 & 5.7 \\
    NGC~2835-S & 12.9 & 7.8 & -118.1 & -67.7 & 87.7 & 148.0 & 77.4 & 99.8 & 93.3 & 84.0 & 87.6 & 102.3 & 120.9 & 6.4 \\
    NGC~3351 & 14.5 & 3.9 & -103.0 & -60.1 & 71.0 & 121.2 & 65.3 & 86.5 & 79.9 & 71.7 & 74.7 & 90.0 & 105.7 & 6.3 \\
    NGC~3627 & 39.3 & 24.8 & -86.4 & -42.6 & 103.2 & 186.3 & 71.4 & 86.1 & 81.6 & 72.7 & 75.9 & 92.6 & 109.0 & 7.1 \\
    NGC~4254 & 29.2 & 19.7 & -112.0 & -60.0 & 108.5 & 186.2 & 83.2 & 101.3 & 95.5 & 81.6 & 87.8 & 109.0 & 128.7 & 9.2 \\
    NGC~4303 & 38.2 & 21.1 & -101.2 & -52.5 & 105.2 & 188.3 & 77.4 & 94.2 & 87.6 & 77.8 & 81.4 & 98.1 & 115.6 & 7.3 \\
    NGC~4321 & 20.5 & 12.7 & -86.4 & -46.8 & 77.6 & 134.2 & 61.6 & 77.5 & 72.3 & 64.1 & 67.1 & 80.9 & 97.4 & 5.9 \\
    NGC~4535 & 19.5 & 12.5 & -98.8 & -54.6 & 82.7 & 134.6 & 68.5 & 90.5 & 84.9 & 76.0 & 79.3 & 93.9 & 108.3 & 6.6 \\
    NGC~5068 & 19.5 & 13.5 & -103.2 & -56.2 & 88.0 & 147.2 & 71.5 & 90.5 & 88.2 & 62.5 & 80.3 & 96.9 & 112.1 & 7.2 \\
    NGC~7496 & 14.5 & 5.8 & -81.5 & -46.6 & 61.3 & 104.0 & 53.8 & 71.1 & 65.6 & 57.9 & 60.7 & 73.5 & 87.2 & 5.6 \\
\hline
\end{tabular}
\end{table*}

\subsection{Post-processing}\label{sec_finalclean}

As the final part of our data reduction pipeline, we post-process the images to remove remaining artifacts such as cosmic ray residuals, strong negative features caused by over-subtraction of the continuum level, and foreground stars. 
These post-processing steps provide an optimal ``artifact free'' image set, although some biases may be introduced in the noise.

\subsubsection{Cosmic Ray Correction}
\label{subsec_cosmicrays}

After the main image processing steps, a number of bright cosmic ray residuals not associated with any emission-line source remain in the HST H$\alpha$ images. These are particularly noticeable in the archival datasets that have fewer exposures, and, hence, have less effective cosmic ray removal during the image reduction process (e.g., NGC\,628).

To remove these contaminants we apply the {\sc deepCR} (Deep Learning Based Cosmic Ray Removal for Astronomical Images) package 
\citep{Zhang20} with the ACS-WFC-F606W.pth learned model to generate a probability map of cosmic ray residuals. We apply a conservative 0.9 probability threshold for galaxies observed as part of our PHANGS-HST-H$\alpha$ Treasury survey (GO-17126), and a less restrictive 0.25 threshold for the archival images. Masked pixel values are interpolated and replaced using the {\sc astropy.convolution.interpolate$\_$replace$\_$nans} function.\footnote{In practice, we iterate three times the interpolating function with a kernal with a standard deviation of one pixel.}

\subsubsection{Continuum Over-Subtraction}
\label{subsubsec_oversub}

In a number of cases, strong continuum sources including bright clusters, foreground stars, and even chance supernovae (e.g.\ the case of SN2020oi in NGC\,4321) are over-subtracted and result in negative values in the flux-calibrated, continuum-subtracted HST-H$\alpha$ images. To identify and mitigate these regions on the maps, we create two masks with the locations of negative pixels: a `strong' negative mask which includes pixels with values $<-5\sigma_\mathrm{RMS}$, and another with a lower threshold of $<$$-\sigma_\mathrm{RMS}$. We grow the strong negative mask into the lower threshold mask, and replace values within this final mask with an interpolated value from the {\sc astropy.convolution.interpolate$\_$replace$\_$nans} function. 

\subsubsection{Foreground Star Removal}
\label{subsubsec_stars}

There are a number of bright, saturated foreground stars that are very evident and produce clear artifacts in our HST-H$\alpha$ images. To remove these features, we use previously defined star masks created for the MUSE dataset \citep{Groves23}. These are based on Gaia objects within the MUSE field-of-view, and then differentiated between stars and clusters (galaxy centers are often included) by the presence of the Calcium Triplet and/or abnormal fits. In addition to these, we manually inspected all of the background-subtracted images, and produced an additional catalog of bright foreground stars that result in artifacts. We then substitute regions from both these catalogs with noise profiles that mimic the image as a whole.

\subsubsection{Mosaic Image for NGC 628}
\label{subsubsec_ngc628mosaic}

NGC\,628 was observed using both the WFC3/UVIS (NGC\,628-E) and ACS/WFC (NGC\,628-C) 
instruments (see Tab.\,\ref{tab:galaxysample}). Because these datasets require separate calibration steps (due to differences in filter response and noise properties), we did not merge them into a single mosaic at the initial narrow-band data processing stage. Instead, both images were fully processed individually (i.e., background subtraction, extinction correction, and continuum subtraction), to produce separate continuum-subtracted H$\alpha$ images. With these H$\alpha$ images (and their corresponding error maps) now in equivalent flux units, we combined them into one final H$\alpha$ mosaic. NGC\,628 is the only case in our 19 galaxy sample where this post-processing combination step is required.

\section{H$\alpha$ Nebulae Catalogs and Scientific Applications}
\label{sec:tour}
\label{sec:dataproducts}

\subsection{Final H$\alpha$ Images}\label{sec:finalimages}


 HST-H$\alpha$ emission, seen as the pinkish-red emission in Figure~\ref{fig:pressrelease},
 shows the variety of star-forming environments encompassed by the PHANGS-HST-H$\alpha$ sample.
There are several galaxies with compact star-forming rings  (including NGC~1512, NGC~1672, and NGC~4321).  A number of the spirals in our sample have nuclear star formation, with clumpy H$\alpha$ associated with nuclear star-forming regions (e.g., NGC~1385, NGC~5068, and NGC~7496). In the barred galaxies (e.g., NGC~1512, NGC~1672, NGC~3627, NGC~7496) it is very obvious that star formation does not occur in the bar, but rather in the center and at the ends of the bar \citep[e.g.,][]{FraserMcKelvie20}. Our sample also includes galaxies with flocculent and grand-design spiral arms. In some galaxies, weak inter-arm H$\alpha$ emission is observed, where weak star formation is taking place (e.g., NGC~4254).

The new maps also show a wide range of source morphologies on smaller scales. There are many individual compact and unresolved HII regions, as well as large complex ones.  There are superbubbles \citep{Barnes2023,Watkins23} and partial shells with a range of sizes, and diffuse H$\alpha$ emission as well.

\subsection{Data Products}

The flux-calibrated emission line PHANGS-HST-H$\alpha$ maps will support a number of science projects.
The current data release includes the reduced drizzled images and the flux-calibrated, continuum-subtracted H$\alpha$ maps (more below), which are available via MAST\footnote{\url{https://archive.stsci.edu/hlsp/phangs.html}}, as well as at the Canadian Astronomy Data Centre as part of the PHANGS archive\footnote{\url{https://www.canfar.net/storage/vault/list/phangs/RELEASES}}. 

\begin{itemize}

\item  We release reduced, drizzled images in units of flux density,
which have been aligned and drizzled onto the same pixel grid as the five broad-band PHANGS-HST images from \citet{Lee2022}.  These images include both line and stellar continuum emission and are ideal for including in spectral energy distribution fitting to determine the age and reddening of star clusters ({\bf see \S~\ref{sec:ages}}). For these, we will release the DRC FITS images of the individual pointings as well as mosaicked DRC FITS images for the galaxies which have multiple pointings.

\item  We also release the flux-calibrated, continuum subtracted H$\alpha$ maps in units of erg/s/cm/arcsec$^2$,
that were created as described in Sections~\ref{subsec_fluxan1} through \ref{subsec_errormaps}; images which have gone through the additional post-processing steps are available upon request.
These images  are ideal for identifying individual HII regions, even very small and faint ones which are not detected in the ground-based MUSE/IFU observations ({\bf \S 3.2)}.  They also provide a unique opportunity to measure the radii of HII regions ({\bf \S \ref{sec:radiushii}}), which is quite challenging from the ground and will lead to significantly improved calculations of key stellar feedback processes such as internal pressure from radiation, thermal gas, and winds.  These images also reveal diffuse ionizing gas in spiral galaxies, including the warm ionized gas leaking out of HII regions ({\bf \S \ref{sec:leakage}}).

\end{itemize}






\subsection{Improving Cluster Age, Reddening, and Mass Estimates by Including H$\alpha$}\label{sec:ages}

Age and mass are two of the most basic properties of a star cluster.  Accurate determinations of the ages and masses of clusters are critical to address many fundamental questions related to stellar feedback and star and cluster formation and evolution.  However, low-metallicity, dust-free ancient clusters are difficult to distinguish from young, dust-rich clusters or from intermediate-age moderately dusty clusters when only broad-band optical observations are used. Breaking this age-reddening-metallicity degeneracy is sufficiently challenging that $\approx80$\% of likely ancient globular clusters were incorrectly dated using five broad-band filters to ages $\lea100$~Myr by the original PHANGS-HST pipeline \citep[e.g.,][]{turner21,Whitmore23,Floyd24},
and a significant fraction of intermediate age $\approx20-500$~Myr clusters were best fit by an extremely young age of 1~Myr and moderate-to-high reddening (e.g., Thilker et al., submitted, Henny et al., submitted).

\begin{figure}
\includegraphics[width=\columnwidth]{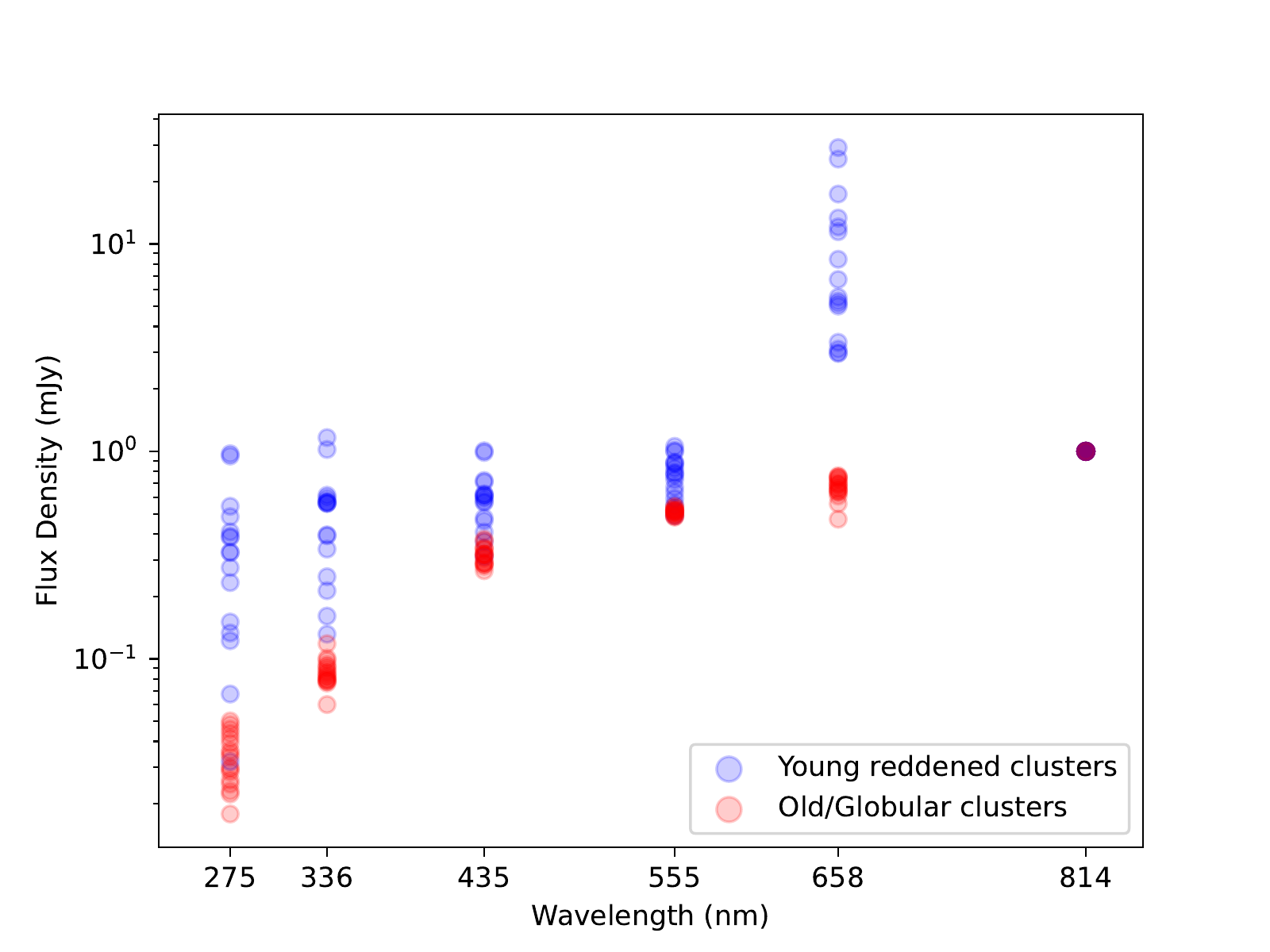}
 \caption{ Median HST-based spectral energy distributions (SEDs) of young, gas and dust-rich clusters (blue points) and old globular clusters (red points) for 10 PHANGS-HST-H$\alpha$ galaxies.  The SEDs are normalized at the I  band. The young and old populations can have similar declining broad-band fluxes towards shorter wavelengths, but have very different H$\alpha$ fluxes (see the clear separation between red and blue data points at 658 nm).  Figure adapted from K. Henny et al., submitted.}
 \label{fig:sed}
\end{figure}

Knowing whether H$\alpha$ line emission is present or absent can significantly improve cluster age estimates \citep[e.g.,][]{chandar16,Ashworth17,Whitmore23}.
While including a short-wavelength broad-band filter (such as in the near-ultraviolet and/or U band) is important, the line emission measured in a narrow-band filter, like the F658N or F657N filters used in this program, is one of the most direct ways to distinguish very young ($\lea 6$~Myr), gas-rich clusters from older, gas-free clusters.  
Figure~\ref{fig:sed} shows median spectral energy distributions (SEDs) for samples of very young, gas-and-dust rich clusters (shown in blue) and for ancient, dust-free clusters (shown in red) in 10 PHANGS-HST-H$\alpha$ galaxies (K. Henny et al., submitted).  While dust-rich young and dust-free old clusters can have red broad-band optical colors (decreasing flux density at shorter wavelengths), the young clusters have strong H$\alpha$ line emission that clearly distinguishes them from the older clusters.
Improved age, reddening, and mass estimates from SED fitting which includes H$\alpha$ photometry from the images presented here will be presented in Thilker et al. (submitted) and Henny et al., (submitted).

For some clusters with no H$\alpha$ emission, it can still be challenging to separate those with ages $\approx7-10$~Myr and moderate reddening from somewhat older, $\approx20-50$~Myr clusters with little reddening.  In these cases, the presence or absence of near-infrared emission from red supergiants with ages $\approx7-10$~Myr in NIRCAM photometry further improves age dating (K. Henny et al., submitted).


\subsection{New Catalog of Ionized Gas Nebulae}

Warm ionized gas in spiral galaxies originates from HII regions, supernovae/SNRs, and planetary nebulae.  Ultraviolet photons can escape from these nebulae and ionize gas far from their birth sites, giving rise to diffuse ionized gas (DIG). Identifying nebulae, particularly compact ones, and separating them from DIG in nearby galaxies can be challenging in imaging and IFU/spectroscopic data with resolution on the order of $1\arcsec$, typical of ground-based observations.

In star-forming galaxies, ionized gas nebulae have a larger range in sizes and more complex morphologies than star clusters.  Panels (a) through (e) in Figure~\ref{fig:HIIcounts} show an example of a large ionized gas region in NGC~1672.  The left panel (a) highlights two dominant star clusters and the stellar population in a $\sim$kpc section of NGC~1672 in a composite BVI image from the PHANGS-HST survey (Lee et al. 2022). Panel~(b) shows H$\alpha$ contours in this region from the PHANGS-MUSE map, and identifies the level of the DIG. 
The proximity of the DIG to HII regions in these images indicates this gas is comparable to the DIG observed in the Milky Way using radio recombination lines which trace denser, ionized gas (e.g. GDIGS survey, \citealt{Anderson21}). 
This is distinct from more diffuse gas, commonly referred to as the Warm Ionized Medium (WIM) that is seen with H-alpha emission in the Milky
Way with the Wisconsin H$\alpha$ Mapper (WHAM) \citep[][]{Haffner03, Krishnarao19}.
Panel~(c) shows the HST continuum-subtracted H$\alpha$, with the nebulae defined by MUSE indicated by the white contours. Panels (d) and (e) show that with high-resolution HST-H$\alpha$ images we can dissect complex nebulae and identify the stellar sources (clusters and massive stars, shown as the blue stars) that produce ionizing photons and drive the larger and smaller shells and bubbles (represented by the black curved lines). Panel~(f) highlights the range of shell-like structures possible even in more compact nebulae, showing H$\alpha$ in the left panel of each image pair, and BVH$\alpha$ in the right panel.

Different methods can be used to identify nebulae.  We will present a new catalog of thousands of ionized gas nebulae in an accompanying work (paper II, Barnes et al., in prep) based on the PHANGS-HST-H$\alpha$ flux-calibrated images described in Section~4. We start from the PHANGS-MUSE nebulae catalogs produced by Groves et al. (2023) in order 
to include flux measurements from the PHANGS-MUSE IFU spectra of key emission lines (including H$\alpha$, H$\beta$, [OIII], and [NII]).

\subsection{Sizes of HII Regions and Constraints on Pre-Supernova Feedback}\label{sec:radiushii}

Assuming the expansion of an HII region from a point source into a homogeneous medium, the pressure $P$ at the inner working surface of the bubble walls scales with radius $r$ as follows: (1) $P_{\rm gas} \propto r^{-3/2}$ for the thermal pressure exerted by the  warm ionized gas that fills the bubble cavity, (2) $P_{\rm rad} \propto r^{-2}$ for the intense radiation field produced by the massive stars in the cluster, and similarly (3) $P_{\rm wind} \propto r^{-2}$ for the mechanical pressure from strong stellar winds. 
(See \citealt[]{Tielens2010,Draine2011,klessen2016,rahner2017} for more details).

\begin{figure}.

\includegraphics[width=\columnwidth]{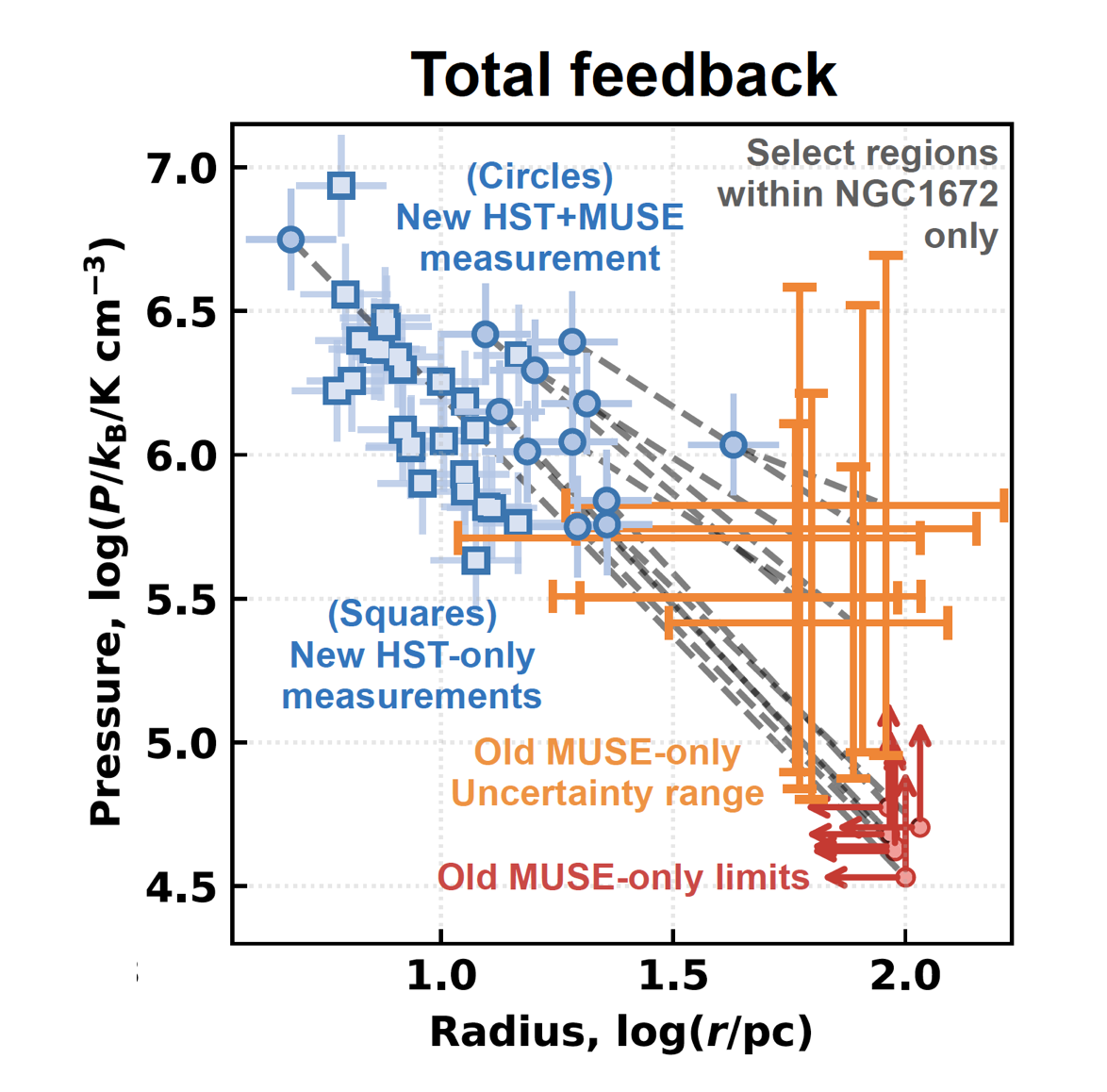}
 \caption{Demonstration of how measurements of the radii of HII regions from HST H$\alpha$ images drastically reduce the uncertainties on stellar feedback calculations.  
 The red points and orange bars indicate the range in size and pressure spanned by the lower and upper limits determined for HII regions in NGC~1672 \citep{Barnes2021}. The dominant uncertainty comes from having only upper limits on HII region size measurements in the ground-based observations. The blue points show that with the addition of HST-based size measurements, the pressure can be measured to within a factor of $\sim2-3$, rather than being constrained to a factor of $\sim100$ . 
  }\label{fig:radP}
\end{figure}

Previously, \citet{Barnes2021}constrained these internal pressure terms in over 5000 HII~ regions using ground-based MUSE/IFU spectroscopy.  However, the best ground-based observations only provide upper limits on the radii of $\sim90$\% of the HII regions in our sample galaxies, since they are unresolved at the $\sim50-100$~pc resolution provided by available ground-based observations. With only weak upper limits on their radii, the pressure calculations have large uncertainties, of about two orders of magnitude, as shown by the red and orange bars in Figure~\ref{fig:radP}.

PHANGS-HST-H$\alpha$ observations allow us to directly measure the radii of HII regions, and also the mass and age of the ionizing cluster. 
Direct size measurements decrease the uncertainties in the pressure estimates from a factor of $\approx100$ to $\approx2-3$,
as shown by the blue data points in Figure~\ref{fig:radP}).
The new calculations will allow the dominant source of feedback to be identified in large samples of 
young clusters as a function of mass and age and across different galactic environments. The method for measuring HII region sizes will be presented in paper~II (Barnes et al., in prep).

\subsection{Leakage of Ionizing Photons from HII Regions}\label{sec:leakage}

\begin{figure*}[!htbp]
\includegraphics[width=\textwidth]{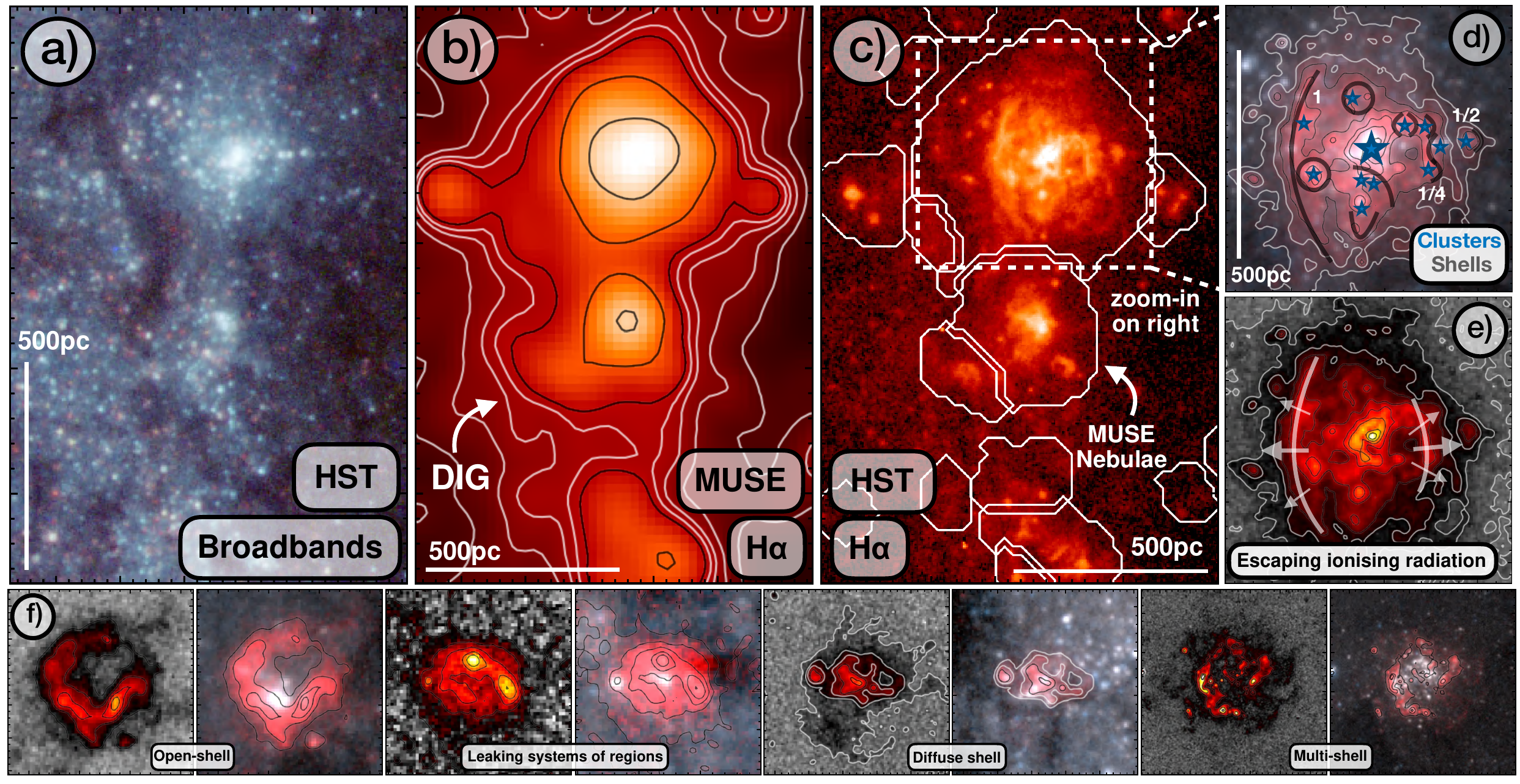}
 \caption{A star-forming complex in NGC~1672 is shown in optical broad-band HST filters in panel~(a) and in HST-H$\alpha$ in panel~(c).  MUSE-defined HII regions and diffuse ionized gas (DIG) are outlined in panel~(b).  
 The complexity of the HII region is highlighted in Panel~(d), which shows the locations of several star clusters and massive stars (star symbols) and full to partial shells of ionized gas (curved lines).  Ionizing radiation leaks out of HII regions from open shells and beyond their walls, as illustrated in panel~(e). Panel~(f) shows examples of different kinds of shell structures visible even in compact nebulae, many of which are leaking ionizing radiation into their surroundings.  The left image in each pair shows just H$\alpha$ emission, and the right image is a composite of B, V, and H$\alpha$.} 
 \label{fig:HIIcounts}
\end{figure*}

Ionizing photons leak out from HII regions and produce diffuse ionized emission (DIG) far from their birth sites. The PHANGS-HST-H$\alpha$ observations are ideal for directly measuring the amount of ionizing radiation leaking out of HII regions, $f_{\rm esc}$, right at the source. 
In Figure~\ref{fig:HIIcounts} panels~(e) and (f), we show examples of diffuse H$\alpha$ emission escaping from HII regions with a variety of structures, some with open shells and some not.
Measurements of H$\alpha$ flux outside of HII~regions (for example along the outer curved shell wall as shown in panel~e)
can be compared with that measured inside HII regions to quantify $f_{\rm esc}$ starting at very early ages.

These local measurements of $f_{\rm esc}$ and HII region morphology can be used to test predictions of how $f_{\rm esc}$ depends on the molecular gas density of the local environment, which will be measured from the ALMA CO(2-1) maps.
For example, simulations  (Kim et al. 2019) indicate that H II regions which form in clouds with high surface density spend a longer time (relative to the free-fall time) in the embedded phase, during which the escape fraction of ionizing radiation is predicted to be small. Simulations  also predict a strong (anti-)correlation of the instantaneous $f_{\rm esc}$ with extinction (which we will estimate from the MUSE-based Balmer decrement), and show that $f_{\rm esc}$ sensitively probes gas/dust substructure driven by turbulence that creates channels for photons to emerge.  The high resolution, near-infrared images from JWST for this same sample of galaxies will also shed light on the presence of holes vs. dense gas.

\section{Summary}
\label{sec:summary}

We have presented PHANGS-HST-H$\alpha$, a Cycle 30 Treasury program which forms an important part of the observational data sets gathered by the PHANGS collaboration (Schinnerer et al. 2019). PHANGS-HST-H$\alpha$ obtained images of 19 nearby ($\lea20$~Mpc) spiral galaxies in the F658N or F657N narrow-band filter with the WFC3 camera.

Our flux calibration, continuum-subtraction, and final cleaning procedures were described in detail in Section~\ref{sec_fluxan}, and are made available online in the \texttt{PyHSTHAContSub} pipeline.
Data products including the drizzled, aligned narrow-band images as well as the flux-calibrated, continuum-subtracted H$\alpha$ maps will be available through MAST and CADC.
The error maps associated with the final continuum-subtracted H$\alpha$ images will also be available. A high-level summary of the steps to create the final flux-calibrated H$\alpha$ and associated error maps is given below.

\begin{enumerate}
  \item \textbf{Create synthetic MUSE images (\S\,\ref{subsec_fluxan1}):} 
  We created new, synthetic images from the MUSE observations to use for flux calibrating the HST-H$\alpha$ images. These were generated to mimic the narrow-band F657N and F658N HST (ACS and WFC3) filters (which contain the H$\alpha$ emission) and the broad-band F555W and F814W (ACS and WFC3) filters (which are used to determine the continuum level).\smallskip
 
  \item \textbf{Convert to flux density units (\S\,\ref{subsec_fluxan2}):}
  All HST and synthetic MUSE images were converted to flux density units (erg/cm\textsuperscript{2}/s/\AA/pixel). \smallskip

  \item \textbf{Smooth and regrid the HST images (\S\,\ref{subsec_fluxan3}):} 
  We smoothed and regridded the narrow- and broad-band HST images to match the resolution and pixel grid of the synthetic MUSE images.\smallskip

  \item \textbf{Determine background flux level from MUSE observations (\S\,\ref{subsec_fluxan4}):} We performed a best linear fit between (smoothed) HST and synthetic MUSE pixel fluxes for each filter, and applied the zero-point offset from these fits to the full resolution HST images. 

  \item \textbf{Continuum Subtraction Algorithm (\S\,\ref{subsec_contsub1}):}
  We calculated the contribution from the stellar continuum in our HST-H$\alpha$ maps from the weighted pixel fluxes in the F555W and F814W images (where the weights (\(W\)) are calculated from the relative wavelength differences).  The continuum image was subtracted from the HST-H$\alpha$ image, and the units converted to flux using the width of the H$\alpha$ filter to produce the final continuum subtracted H$\alpha$ image (\(F_\mathrm{H\alpha}\)).

  \item \textbf{Correction for [N\,II] Emission (\S\,\ref{subsec_contsub2}):}
  In addition to H$\alpha$, the F657N and F658N filters include some emission from the [N\,II]\,$\lambda$6548\AA\ and [N\,II]\,$\lambda$6583\AA\ emission lines. We used the MUSE spectra to determine a single correction factor for [N\,II] emission for each galaxy. After this step, our pipeline produced a continuum-subtracted, H$\alpha$-only emission line map for each galaxy.

  \item \textbf{Error Maps (\S\,\ref{subsec_errormaps}):}
  We produced error maps associated with the reduced HST-H$\alpha$ images by propagating errors in the individual pixels from the narrow- and broad-band images.

  \item \textbf{Final Post-processing Steps (\S\,\ref{sec_finalclean}):}
  These steps included corrections for remaining cosmic rays, continuum over-subtraction, and the removal of foreground stars. We also produce a final \Ha\ mosaic image for NGC\,628, since the two pointings were observed with different instruments.

\end{enumerate}

In Section 5 we discussed several of the key science goals of the PHANGS-HST-H$\alpha$ survey, which include: improved age-dating of star clusters, size measurements of HII regions and calculations of pre-supernova pressure (from warm thermal gas, radiation, and stellar winds), and direct measurements of ionizing photons leaking out of HII regions. Catalogs of nebulae will be produced from our continuum-subtracted, flux calibrated HST-H$\alpha$ maps and released as part of an accompanying paper (paper II, Barnes et al.)

\section*{Acknowledgements}

This research has made use of the NASA/IPAC Extragalactic Database (NED) which is operated by the Jet Propulsion Laboratory, California Institute of Technology, under contract with NASA. 

The {\it HST} images produced in this paper can be found in MAST with DOI: \dataset[10.17909/t9-r08f-dq31]{https://archive.stsci.edu/hlsp/phangs/phangs-hst}

MB gratefully acknowledges support from the ANID BASAL project FB210003 and from the FONDECYT regular grant 1211000. This work was supported by the French government through the France 2030 investment plan managed by the National Research Agency (ANR), as part of the Initiative of Excellence of Université Côte d'Azur under reference number ANR-15-IDEX-01.
OE acknowledges funding from the Deutsche Forschungsgemeinschaft (DFG, German Research Foundation) -- project-ID 541068876.
SCOG and RSK acknowledge financial support from the European Research Council via the ERC Synergy Grant ``ECOGAL'' (project ID 855130),  from the German Excellence Strategy via the Heidelberg Cluster of Excellence (EXC 2181 - 390900948) ``STRUCTURES'', and from the German Ministry for Economic Affairs and Climate Action in project ``MAINN'' (funding ID 50OO2206). RSK is also grateful for computing resources provided by the Ministry of Science, Research and the Arts (MWK) of the State of Baden-W\"{u}rttemberg through bwHPC and the German Science Foundation (DFG) through grants INST 35/1134-1 FUGG and 35/1597-1 FUGG, and also for data storage at SDS@hd funded through grants INST 35/1314-1 FUGG and INST 35/1503-1 FUGG. RSK also thanks the Harvard-Smithsonian Center for Astrophysics and the Radcliffe Institute for Advanced Studies for their hospitality during his sabbatical, and the 2024/25 Class of Radcliffe Fellows for highly interesting and stimulating discussions. 
KK gratefully acknowledges funding from the Deutsche Forschungsgemeinschaft (DFG, German Research Foundation) in the form of an Emmy Noether Research Group (grant number KR4598/2-1, PI Kreckel) and the European Research Council’s starting grant ERC StG-101077573 (``ISM-METALS"). G.A.B. acknowledges the support from the ANID Basal project FB210003.
JS acknowledges support by the National Aeronautics and Space Administration (NASA) through the NASA Hubble Fellowship grant HST-HF2-51544 awarded by the Space Telescope Science Institute (STScI), operated by the Association of Universities for Research in Astronomy, Inc., under contract NAS~5-26555.

We thank the anonymous referee for a quick and helpful report.

\software{PyRAF \citep{pyraf}, Astrodrizzle \citep{astrodrizzle}, DOLPHOT \citep[v2.0][]{dolphot}, Photutils \citep{photutils}, CIGALE \citep{burgarella05,noll09,boquien19}}

\bibliographystyle{aasjournal}
\bibliography{references_all,references_deep, references_code} 

\begin{thebibliography}{}
\expandafter\ifx\csname natexlab\endcsname\relax\def\natexlab#1{#1}\fi
\providecommand{\url}[1]{\href{#1}{#1}}
\providecommand{\dodoi}[1]{doi:~\href{http://doi.org/#1}{\nolinkurl{#1}}}
\providecommand{\doeprint}[1]{\href{http://ascl.net/#1}{\nolinkurl{http://ascl.net/#1}}}
\providecommand{\doarXiv}[1]{\href{https://arxiv.org/abs/#1}{\nolinkurl{https://arxiv.org/abs/#1}}}

\bibitem[{{Anand} {et~al.}(2021){Anand}, {Lee}, {Van Dyk}, {Leroy}, {Rosolowsky}, {Schinnerer}, {Larson}, {Kourkchi}, {Kreckel}, {Scheuermann}, {Rizzi}, {Thilker}, {Tully}, {Bigiel}, {Blanc}, {Boquien}, {Chandar}, {Dale}, {Emsellem}, {Deger}, {Glover}, {Grasha}, {Groves}, {S. Klessen}, {Kruijssen}, {Querejeta}, {S{\'a}nchez-Bl{\'a}zquez}, {Schruba}, {Turner}, {Ubeda}, {Williams}, \& {Whitmore}}]{Anand21}
{Anand}, G.~S., {Lee}, J.~C., {Van Dyk}, S.~D., {et~al.} 2021, \mnras, 501, 3621, \dodoi{10.1093/mnras/staa3668}

\bibitem[{{Anderson} {et~al.}(2014){Anderson}, {Bania}, {Balser}, {Cunningham}, {Wenger}, {Johnstone}, \& {Armentrout}}]{Anderson14}
{Anderson}, L.~D., {Bania}, T.~M., {Balser}, D.~S., {et~al.} 2014, \apjs, 212, 1, \dodoi{10.1088/0067-0049/212/1/1}

\bibitem[{{Anderson} {et~al.}(2021){Anderson}, {Luisi}, {Liu}, {Wenger}, {Balser}, {Bania}, {Haffner}, {Linville}, \& {Mascoop}}]{Anderson21}
{Anderson}, L.~D., {Luisi}, M., {Liu}, B., {et~al.} 2021, \apjs, 254, 28, \dodoi{10.3847/1538-4365/abef65}

\bibitem[{{Ashworth} {et~al.}(2017){Ashworth}, {Fumagalli}, {Krumholz}, {Adamo}, {Calzetti}, {Chandar}, {Cignoni}, {Dale}, {Elmegreen}, {Gallagher}, {Gouliermis}, {Grasha}, {Grebel}, {Johnson}, {Lee}, {Tosi}, \& {Wofford}}]{Ashworth17}
{Ashworth}, G., {Fumagalli}, M., {Krumholz}, M.~R., {et~al.} 2017, \mnras, 469, 2464, \dodoi{10.1093/mnras/stx935}

\bibitem[{{Astropy Collaboration} {et~al.}(2018){Astropy Collaboration}, {Price-Whelan}, {Sip{\H{o}}cz}, {G{\"u}nther}, {Lim}, {Crawford}, {Conseil}, {Shupe}, {Craig}, {Dencheva}, {Ginsburg}, {VanderPlas}, {Bradley}, {P{\'e}rez-Su{\'a}rez}, {de Val-Borro}, {Aldcroft}, {Cruz}, {Robitaille}, {Tollerud}, {Ardelean}, {Babej}, {Bach}, {Bachetti}, {Bakanov}, {Bamford}, {Barentsen}, {Barmby}, {Baumbach}, {Berry}, {Biscani}, {Boquien}, {Bostroem}, {Bouma}, {Brammer}, {Bray}, {Breytenbach}, {Buddelmeijer}, {Burke}, {Calderone}, {Cano Rodr{\'\i}guez}, {Cara}, {Cardoso}, {Cheedella}, {Copin}, {Corrales}, {Crichton}, {D'Avella}, {Deil}, {Depagne}, {Dietrich}, {Donath}, {Droettboom}, {Earl}, {Erben}, {Fabbro}, {Ferreira}, {Finethy}, {Fox}, {Garrison}, {Gibbons}, {Goldstein}, {Gommers}, {Greco}, {Greenfield}, {Groener}, {Grollier}, {Hagen}, {Hirst}, {Homeier}, {Horton}, {Hosseinzadeh}, {Hu}, {Hunkeler}, {Ivezi{\'c}}, {Jain}, {Jenness}, {Kanarek}, {Kendrew}, {Kern}, {Kerzendorf}, {Khvalko}, {King}, {Kirkby}, {Kulkarni},
  {Kumar}, {Lee}, {Lenz}, {Littlefair}, {Ma}, {Macleod}, {Mastropietro}, {McCully}, {Montagnac}, {Morris}, {Mueller}, {Mumford}, {Muna}, {Murphy}, {Nelson}, {Nguyen}, {Ninan}, {N{\"o}the}, {Ogaz}, {Oh}, {Parejko}, {Parley}, {Pascual}, {Patil}, {Patil}, {Plunkett}, {Prochaska}, {Rastogi}, {Reddy Janga}, {Sabater}, {Sakurikar}, {Seifert}, {Sherbert}, {Sherwood-Taylor}, {Shih}, {Sick}, {Silbiger}, {Singanamalla}, {Singer}, {Sladen}, {Sooley}, {Sornarajah}, {Streicher}, {Teuben}, {Thomas}, {Tremblay}, {Turner}, {Terr{\'o}n}, {van Kerkwijk}, {de la Vega}, {Watkins}, {Weaver}, {Whitmore}, {Woillez}, {Zabalza}, \& {Astropy Contributors}}]{AstropyCollaboration2018}
{Astropy Collaboration}, {Price-Whelan}, A.~M., {Sip{\H{o}}cz}, B.~M., {et~al.} 2018, \aj, 156, 123, \dodoi{10.3847/1538-3881/aabc4f}

\bibitem[{Barnes(2025)}]{ashley_barnes_2025_14610187}
Barnes, A. 2025, ashleythomasbarnes/reduction\_phangs\_hst: Version 1 release of the Phangs HST continuum subtraction pipeline, v1.0.0,  Zenodo, \dodoi{10.5281/zenodo.14610187}

\bibitem[{{Barnes} {et~al.}(2021){Barnes}, {Glover}, {Kreckel}, {Ostriker}, {Bigiel}, {Belfiore}, {Be{\v{s}}li{\'c}}, {Blanc}, {Chevance}, {Dale}, {Egorov}, {Eibensteiner}, {Emsellem}, {Grasha}, {Groves}, {Klessen}, {Kruijssen}, {Leroy}, {Longmore}, {Lopez}, {McElroy}, {Meidt}, {Murphy}, {Rosolowsky}, {Saito}, {Santoro}, {Schinnerer}, {Schruba}, {Sun}, {Watkins}, \& {Williams}}]{Barnes2021}
{Barnes}, A.~T., {Glover}, S.~C.~O., {Kreckel}, K., {et~al.} 2021, \mnras, 508, 5362, \dodoi{10.1093/mnras/stab2958}

\bibitem[{{Barnes} {et~al.}(2023){Barnes}, {Watkins}, {Meidt}, {Kreckel}, {Sormani}, {Tre{\ss}}, {Glover}, {Bigiel}, {Chandar}, {Emsellem}, {Lee}, {Leroy}, {Sandstrom}, {Schinnerer}, {Rosolowsky}, {Belfiore}, {Blanc}, {Boquien}, {Brok}, {Cao}, {Chevance}, {Dale}, {Egorov}, {Eibensteiner}, {Grasha}, {Groves}, {Hassani}, {Henshaw}, {Jeffreson}, {Jim{\'e}nez-Donaire}, {Keller}, {Klessen}, {Koch}, {Kruijssen}, {Larson}, {Li}, {Liu}, {Lopez}, {Murphy}, {Neumann}, {Pety}, {Pinna}, {Querejeta}, {Renaud}, {Saito}, {Sarbadhicary}, {Sardone}, {Smith}, {Stuber}, {Sun}, {Thilker}, {Usero}, {Whitmore}, \& {Williams}}]{Barnes2023}
{Barnes}, A.~T., {Watkins}, E.~J., {Meidt}, S.~E., {et~al.} 2023, \apjl, 944, L22, \dodoi{10.3847/2041-8213/aca7b9}

\bibitem[{{Belfiore} {et~al.}(2022){Belfiore}, {Santoro}, {Groves}, {Schinnerer}, {Kreckel}, {Glover}, {Klessen}, {Emsellem}, {Blanc}, {Congiu}, {Barnes}, {Boquien}, {Chevance}, {Dale}, {Kruijssen}, {Leroy}, {Pan}, {Pessa}, {Schruba}, \& {Williams}}]{Belfiore2022}
{Belfiore}, F., {Santoro}, F., {Groves}, B., {et~al.} 2022, \aap, 659, A26, \dodoi{10.1051/0004-6361/202141859}

\bibitem[{{Boquien} {et~al.}(2019){Boquien}, {Burgarella}, {Roehlly}, {Buat}, {Ciesla}, {Corre}, {Inoue}, \& {Salas}}]{boquien19}
{Boquien}, M., {Burgarella}, D., {Roehlly}, Y., {et~al.} 2019, \aap, 622, A103, \dodoi{10.1051/0004-6361/201834156}

\bibitem[{{Bradley} {et~al.}(2019){Bradley}, {Sipocz}, {Robitaille}, {Tollerud}, {Vin{\'\i}cius}, {Deil}, {Barbary}, {G{\"u}nther}, {Cara}, {Busko}, {Conseil}, {Droettboom}, {Bostroem}, {Bray}, {Andersen Bratholm}, {Wilson}, {Craig}, {Barentsen}, {Pascual}, {Donath}, {Greco}, {Perren}, {Lim}, \& {Kerzendorf}}]{photutils}
{Bradley}, L., {Sipocz}, B., {Robitaille}, T., {et~al.} 2019, {astropy/photutils: v0.6}, v0.6,  Zenodo, \dodoi{10.5281/zenodo.2533376}

\bibitem[{{Brinchmann} {et~al.}(2004){Brinchmann}, {Charlot}, {White}, {Tremonti}, {Kauffmann}, {Heckman}, \& {Brinkmann}}]{brinchmann04}
{Brinchmann}, J., {Charlot}, S., {White}, S.~D.~M., {et~al.} 2004, \mnras, 351, 1151, \dodoi{10.1111/j.1365-2966.2004.07881.x}

\bibitem[{{Burgarella} {et~al.}(2005){Burgarella}, {Buat}, \& {Iglesias-P{\'a}ramo}}]{burgarella05}
{Burgarella}, D., {Buat}, V., \& {Iglesias-P{\'a}ramo}, J. 2005, \mnras, 360, 1413, \dodoi{10.1111/j.1365-2966.2005.09131.x}

\bibitem[{{Cardelli} {et~al.}(1989){Cardelli}, {Clayton}, \& {Mathis}}]{cardelli89}
{Cardelli}, J.~A., {Clayton}, G.~C., \& {Mathis}, J.~S. 1989, \apj, 345, 245, \dodoi{10.1086/167900}

\bibitem[{{Caswell} \& {Haynes}(1987)}]{Caswell1987}
{Caswell}, J.~L., \& {Haynes}, R.~F. 1987, \aap, 171, 261

\bibitem[{{Chandar} {et~al.}(2016){Chandar}, {Whitmore}, {Dinino}, {Kennicutt}, {Chien}, {Schinnerer}, \& {Meidt}}]{chandar16}
{Chandar}, R., {Whitmore}, B.~C., {Dinino}, D., {et~al.} 2016, \apj, 824, 71, \dodoi{10.3847/0004-637X/824/2/71}

\bibitem[{{Chevance} {et~al.}(2020){Chevance}, {Kruijssen}, {Vazquez-Semadeni}, {Nakamura}, {Klessen}, {Ballesteros-Paredes}, {Inutsuka}, {Adamo}, \& {Hennebelle}}]{chevance2020}
{Chevance}, M., {Kruijssen}, J.~M.~D., {Vazquez-Semadeni}, E., {et~al.} 2020, \ssr, 216, 50, \dodoi{10.1007/s11214-020-00674-x}

\bibitem[{{Churchwell}(2002)}]{Churchwell2002}
{Churchwell}, E. 2002, \araa, 40, 27, \dodoi{10.1146/annurev.astro.40.060401.093845}

\bibitem[{{Courtois} {et~al.}(2012){Courtois}, {Hoffman}, {Tully}, \& {Gottl{\"o}ber}}]{CosmicFlows1}
{Courtois}, H.~M., {Hoffman}, Y., {Tully}, R.~B., \& {Gottl{\"o}ber}, S. 2012, \apj, 744, 43, \dodoi{10.1088/0004-637X/744/1/43}

\bibitem[{{Dolphin}(2016)}]{dolphot}
{Dolphin}, A. 2016, {DOLPHOT: Stellar photometry}.
\newblock \doeprint{1608.013}

\bibitem[{{Draine}(2011)}]{Draine2011}
{Draine}, B.~T. 2011, {Physics of the Interstellar and Intergalactic Medium}

\bibitem[{{Emsellem} {et~al.}(2022){Emsellem}, {Schinnerer}, {Santoro}, {Belfiore}, {Pessa}, {McElroy}, {Blanc}, {Congiu}, {Groves}, {Ho}, {Kreckel}, {Razza}, {Sanchez-Blazquez}, {Egorov}, {Faesi}, {Klessen}, {Leroy}, {Meidt}, {Querejeta}, {Rosolowsky}, {Scheuermann}, {Anand}, {Barnes}, {Be{\v{s}}li{\'c}}, {Bigiel}, {Boquien}, {Cao}, {Chevance}, {Dale}, {Eibensteiner}, {Glover}, {Grasha}, {Henshaw}, {Hughes}, {Koch}, {Kruijssen}, {Lee}, {Liu}, {Pan}, {Pety}, {Saito}, {Sandstrom}, {Schruba}, {Sun}, {Thilker}, {Usero}, {Watkins}, \& {Williams}}]{Emsellem2022}
{Emsellem}, E., {Schinnerer}, E., {Santoro}, F., {et~al.} 2022, \aap, 659, A191, \dodoi{10.1051/0004-6361/202141727}

\bibitem[{{Floyd} {et~al.}(2024){Floyd}, {Chandar}, {Whitmore}, {Thilker}, {Lee}, {Pauline}, {Thomas}, {Berschback}, {Henny}, {Dale}, {Klessen}, {Schinnerer}, {Grasha}, {Boquien}, {Larson}, {Deger}, {Barnes}, {Leroy}, {Rosolowsky}, {Williams}, \& {{\'U}beda}}]{Floyd24}
{Floyd}, M., {Chandar}, R., {Whitmore}, B.~C., {et~al.} 2024, \aj, 167, 95, \dodoi{10.3847/1538-3881/ad1889}

\bibitem[{{Fraser-McKelvie} {et~al.}(2020){Fraser-McKelvie}, {Arag{\'o}n-Salamanca}, {Merrifield}, {Masters}, {Nair}, {Emsellem}, {Kraljic}, {Krishnarao}, {Andrews}, {Drory}, \& {Neumann}}]{FraserMcKelvie20}
{Fraser-McKelvie}, A., {Arag{\'o}n-Salamanca}, A., {Merrifield}, M., {et~al.} 2020, \mnras, 495, 4158, \dodoi{10.1093/mnras/staa1416}

\bibitem[{{Genzel} {et~al.}(2010){Genzel}, {Tacconi}, {Gracia-Carpio}, {Sternberg}, {Cooper}, {Shapiro}, {Bolatto}, {Bouch{\'e}}, {Bournaud}, {Burkert}, {Combes}, {Comerford}, {Cox}, {Davis}, {F{\"o}rster Schreiber}, {Garcia-Burillo}, {Lutz}, {Naab}, {Neri}, {Omont}, {Shapley}, \& {Weiner}}]{Genzel10}
{Genzel}, R., {Tacconi}, L.~J., {Gracia-Carpio}, J., {et~al.} 2010, \mnras, 407, 2091, \dodoi{10.1111/j.1365-2966.2010.16969.x}

\bibitem[{{Groves} {et~al.}(2023){Groves}, {Kreckel}, {Santoro}, {Belfiore}, {Zavodnik}, {Congiu}, {Egorov}, {Emsellem}, {Grasha}, {Leroy}, {Scheuermann}, {Schinnerer}, {Watkins}, {Barnes}, {Bigiel}, {Dale}, {Glover}, {Pessa}, {Sanchez-Blazquez}, \& {Williams}}]{Groves23}
{Groves}, B., {Kreckel}, K., {Santoro}, F., {et~al.} 2023, \mnras, 520, 4902, \dodoi{10.1093/mnras/stad114}

\bibitem[{{Haffner} {et~al.}(2003){Haffner}, {Reynolds}, {Tufte}, {Madsen}, {Jaehnig}, \& {Percival}}]{Haffner03}
{Haffner}, L.~M., {Reynolds}, R.~J., {Tufte}, S.~L., {et~al.} 2003, \apjs, 149, 405, \dodoi{10.1086/378850}

\bibitem[{{Haffner} {et~al.}(2009){Haffner}, {Dettmar}, {Beckman}, {Wood}, {Slavin}, {Giammanco}, {Madsen}, {Zurita}, \& {Reynolds}}]{Haffner09}
{Haffner}, L.~M., {Dettmar}, R.~J., {Beckman}, J.~E., {et~al.} 2009, Reviews of Modern Physics, 81, 969, \dodoi{10.1103/RevModPhys.81.969}

\bibitem[{{Klessen} \& {Glover}(2016)}]{klessen2016}
{Klessen}, R.~S., \& {Glover}, S. C.~O. 2016, Saas-Fee Advanced Course, 43, 85, \dodoi{10.1007/978-3-662-47890-5_2}

\bibitem[{{Krishnarao}(2019)}]{Krishnarao19}
{Krishnarao}, D. 2019, The Journal of Open Source Software, 4, 1940, \dodoi{10.21105/joss.01940}

\bibitem[{{Krist} {et~al.}(2011){Krist}, {Hook}, \& {Stoehr}}]{Krist11}
{Krist}, J.~E., {Hook}, R.~N., \& {Stoehr}, F. 2011, in Society of Photo-Optical Instrumentation Engineers (SPIE) Conference Series, Vol. 8127, Optical Modeling and Performance Predictions V, ed. M.~A. {Kahan}, 81270J, \dodoi{10.1117/12.892762}

\bibitem[{{Lee} {et~al.}(2022){Lee}, {Whitmore}, {Thilker}, {Deger}, {Larson}, {Ubeda}, {Anand}, {Boquien}, {Chandar}, {Dale}, {Emsellem}, {Leroy}, {Rosolowsky}, {Schinnerer}, {Schmidt}, {Lilly}, {Turner}, {Van Dyk}, {White}, {Barnes}, {Belfiore}, {Bigiel}, {Blanc}, {Cao}, {Chevance}, {Congiu}, {Egorov}, {Glover}, {Grasha}, {Groves}, {Henshaw}, {Hughes}, {Klessen}, {Koch}, {Kreckel}, {Kruijssen}, {Liu}, {Lopez}, {Mayker}, {Meidt}, {Murphy}, {Pan}, {Pety}, {Querejeta}, {Razza}, {Saito}, {S{\'a}nchez-Bl{\'a}zquez}, {Santoro}, {Sardone}, {Scheuermann}, {Schruba}, {Sun}, {Usero}, {Watkins}, \& {Williams}}]{Lee2022}
{Lee}, J.~C., {Whitmore}, B.~C., {Thilker}, D.~A., {et~al.} 2022, \apjs, 258, 10, \dodoi{10.3847/1538-4365/ac1fe5}

\bibitem[{{Lee} {et~al.}(2023){Lee}, {Sandstrom}, {Leroy}, {Thilker}, {Schinnerer}, {Rosolowsky}, {Larson}, {Egorov}, {Williams}, {Schmidt}, {Emsellem}, {Anand}, {Barnes}, {Belfiore}, {Be{\v{s}}li{\'c}}, {Bigiel}, {Blanc}, {Bolatto}, {Boquien}, {den Brok}, {Cao}, {Chandar}, {Chastenet}, {Chevance}, {Chiang}, {Congiu}, {Dale}, {Deger}, {Eibensteiner}, {Faesi}, {Glover}, {Grasha}, {Groves}, {Hassani}, {Henny}, {Henshaw}, {Hoyer}, {Hughes}, {Jeffreson}, {Jim{\'e}nez-Donaire}, {Kim}, {Kim}, {Klessen}, {Koch}, {Kreckel}, {Kruijssen}, {Li}, {Liu}, {Lopez}, {Maschmann}, {Chen}, {Meidt}, {Murphy}, {Neumann}, {Neumayer}, {Pan}, {Pessa}, {Pety}, {Querejeta}, {Pinna}, {Rodr{\'\i}guez}, {Saito}, {S{\'a}nchez-Bl{\'a}zquez}, {Santoro}, {Sardone}, {Smith}, {Sormani}, {Scheuermann}, {Stuber}, {Sutter}, {Sun}, {Teng}, {Tre{\ss}}, {Usero}, {Watkins}, {Whitmore}, \& {Razza}}]{Lee2023}
{Lee}, J.~C., {Sandstrom}, K.~M., {Leroy}, A.~K., {et~al.} 2023, \apjl, 944, L17, \dodoi{10.3847/2041-8213/acaaae}

\bibitem[{{Leroy} {et~al.}(2019){Leroy}, {Sandstrom}, {Lang}, {Lewis}, {Salim}, {Behrens}, {Chastenet}, {Chiang}, {Gallagher}, {Kessler}, \& {Utomo}}]{leroy19}
{Leroy}, A.~K., {Sandstrom}, K.~M., {Lang}, D., {et~al.} 2019, \apjs, 244, 24, \dodoi{10.3847/1538-4365/ab3925}

\bibitem[{{Leroy} {et~al.}(2021{\natexlab{a}}){Leroy}, {Schinnerer}, {Hughes}, {Rosolowsky}, {Pety}, {Schruba}, {Usero}, {Blanc}, {Chevance}, {Emsellem}, {Faesi}, {Herrera}, {Liu}, {Meidt}, {Querejeta}, {Saito}, {Sandstrom}, {Sun}, {Williams}, {Anand}, {Barnes}, {Behrens}, {Belfiore}, {Benincasa}, {Be{\v{s}}li{\'c}}, {Bigiel}, {Bolatto}, {den Brok}, {Cao}, {Chandar}, {Chastenet}, {Chiang}, {Congiu}, {Dale}, {Deger}, {Eibensteiner}, {Egorov}, {Garc{\'\i}a-Rodr{\'\i}guez}, {Glover}, {Grasha}, {Henshaw}, {Ho}, {Kepley}, {Kim}, {Klessen}, {Kreckel}, {Koch}, {Kruijssen}, {Larson}, {Lee}, {Lopez}, {Machado}, {Mayker}, {McElroy}, {Murphy}, {Ostriker}, {Pan}, {Pessa}, {Puschnig}, {Razza}, {S{\'a}nchez-Bl{\'a}zquez}, {Santoro}, {Sardone}, {Scheuermann}, {Sliwa}, {Sormani}, {Stuber}, {Thilker}, {Turner}, {Utomo}, {Watkins}, \& {Whitmore}}]{phangs-alma}
{Leroy}, A.~K., {Schinnerer}, E., {Hughes}, A., {et~al.} 2021{\natexlab{a}}, arXiv e-prints, arXiv:2104.07739.
\newblock \doarXiv{2104.07739}

\bibitem[{{Leroy} {et~al.}(2021{\natexlab{b}}){Leroy}, {Schinnerer}, {Hughes}, {Rosolowsky}, {Pety}, {Schruba}, {Usero}, {Blanc}, {Chevance}, {Emsellem}, {Faesi}, {Herrera}, {Liu}, {Meidt}, {Querejeta}, {Saito}, {Sandstrom}, {Sun}, {Williams}, {Anand}, {Barnes}, {Behrens}, {Belfiore}, {Benincasa}, {Be{\v{s}}li{\'c}}, {Bigiel}, {Bolatto}, {den Brok}, {Cao}, {Chandar}, {Chastenet}, {Chiang}, {Congiu}, {Dale}, {Deger}, {Eibensteiner}, {Egorov}, {Garc{\'\i}a-Rodr{\'\i}guez}, {Glover}, {Grasha}, {Henshaw}, {Ho}, {Kepley}, {Kim}, {Klessen}, {Kreckel}, {Koch}, {Kruijssen}, {Larson}, {Lee}, {Lopez}, {Machado}, {Mayker}, {McElroy}, {Murphy}, {Ostriker}, {Pan}, {Pessa}, {Puschnig}, {Razza}, {S{\'a}nchez-Bl{\'a}zquez}, {Santoro}, {Sardone}, {Scheuermann}, {Sliwa}, {Sormani}, {Stuber}, {Thilker}, {Turner}, {Utomo}, {Watkins}, \& {Whitmore}}]{Leroy21}
---. 2021{\natexlab{b}}, \apjs, 257, 43, \dodoi{10.3847/1538-4365/ac17f3}

\bibitem[{{Leroy} {et~al.}(2023){Leroy}, {Sandstrom}, {Rosolowsky}, {Belfiore}, {Bolatto}, {Cao}, {Koch}, {Schinnerer}, {Barnes}, {Be{\v{s}}li{\'c}}, {Bigiel}, {Blanc}, {Chastenet}, {Chen}, {Chevance}, {Chown}, {Congiu}, {Dale}, {Egorov}, {Emsellem}, {Eibensteiner}, {Faesi}, {Glover}, {Grasha}, {Groves}, {Hassani}, {Henshaw}, {Hughes}, {Jim{\'e}nez-Donaire}, {Kim}, {Klessen}, {Kreckel}, {Kruijssen}, {Larson}, {Lee}, {Levy}, {Liu}, {Lopez}, {Meidt}, {Murphy}, {Neumann}, {Pessa}, {Pety}, {Saito}, {Sardone}, {Sun}, {Thilker}, {Usero}, {Watkins}, {Whitcomb}, \& {Williams}}]{Leroy2023}
{Leroy}, A.~K., {Sandstrom}, K., {Rosolowsky}, E., {et~al.} 2023, \apjl, 944, L9, \dodoi{10.3847/2041-8213/acaf85}

\bibitem[{{McLeod} {et~al.}(2020){McLeod}, {Kruijssen}, {Weisz}, {Zeidler}, {Schruba}, {Dalcanton}, {Longmore}, {Chevance}, {Faesi}, \& {Byler}}]{McLeod20}
{McLeod}, A.~F., {Kruijssen}, J.~M.~D., {Weisz}, D.~R., {et~al.} 2020, \apj, 891, 25, \dodoi{10.3847/1538-4357/ab6d63}

\bibitem[{{Noll} {et~al.}(2009){Noll}, {Burgarella}, {Giovannoli}, {Buat}, {Marcillac}, \& {Mu{\~n}oz-Mateos}}]{noll09}
{Noll}, S., {Burgarella}, D., {Giovannoli}, E., {et~al.} 2009, \aap, 507, 1793, \dodoi{10.1051/0004-6361/200912497}

\bibitem[{{Olivier} {et~al.}(2020){Olivier}, {Lopez}, {Rosen}, {Nayak}, {Reiter}, {Krumholz}, \& {Bolatto}}]{olivier21}
{Olivier}, G.~M., {Lopez}, L.~A., {Rosen}, A.~L., {et~al.} 2020, arXiv e-prints, arXiv:2009.10079.
\newblock \doarXiv{2009.10079}

\bibitem[{{Rahner} {et~al.}(2017){Rahner}, {Pellegrini}, {Glover}, \& {Klessen}}]{rahner2017}
{Rahner}, D., {Pellegrini}, E.~W., {Glover}, S. C.~O., \& {Klessen}, R.~S. 2017, \mnras, 470, 4453, \dodoi{10.1093/mnras/stx1532}

\bibitem[{{Saintonge} {et~al.}(2017){Saintonge}, {Catinella}, {Tacconi}, {Kauffmann}, {Genzel}, {Cortese}, {Dav{\'e}}, {Fletcher}, {Graci{\'a}-Carpio}, {Kramer}, {Heckman}, {Janowiecki}, {Lutz}, {Rosario}, {Schiminovich}, {Schuster}, {Wang}, {Wuyts}, {Borthakur}, {Lamperti}, \& {Roberts-Borsani}}]{Saintonge17}
{Saintonge}, A., {Catinella}, B., {Tacconi}, L.~J., {et~al.} 2017, \apjs, 233, 22, \dodoi{10.3847/1538-4365/aa97e0}

\bibitem[{{Salim} {et~al.}(2018){Salim}, {Boquien}, \& {Lee}}]{salim18}
{Salim}, S., {Boquien}, M., \& {Lee}, J.~C. 2018, \apj, 859, 11, \dodoi{10.3847/1538-4357/aabf3c}

\bibitem[{{Salim} {et~al.}(2007){Salim}, {Rich}, {Charlot}, {Brinchmann}, {Johnson}, {Schiminovich}, {Seibert}, {Mallery}, {Heckman}, {Forster}, {Friedman}, {Martin}, {Morrissey}, {Neff}, {Small}, {Wyder}, {Bianchi}, {Donas}, {Lee}, {Madore}, {Milliard}, {Szalay}, {Welsh}, \& {Yi}}]{salim07}
{Salim}, S., {Rich}, R.~M., {Charlot}, S., {et~al.} 2007, \apjs, 173, 267, \dodoi{10.1086/519218}

\bibitem[{{Salim} {et~al.}(2016){Salim}, {Lee}, {Janowiecki}, {da Cunha}, {Dickinson}, {Boquien}, {Burgarella}, {Salzer}, \& {Charlot}}]{salim16}
{Salim}, S., {Lee}, J.~C., {Janowiecki}, S., {et~al.} 2016, \apjs, 227, 2, \dodoi{10.3847/0067-0049/227/1/2}

\bibitem[{{Schlafly} \& {Finkbeiner}(2011)}]{schlafly11}
{Schlafly}, E.~F., \& {Finkbeiner}, D.~P. 2011, \apj, 737, 103, \dodoi{10.1088/0004-637X/737/2/103}

\bibitem[{{Science Software Branch at STScI}(2012)}]{pyraf}
{Science Software Branch at STScI}. 2012, {PyRAF: Python alternative for IRAF}.
\newblock \doeprint{1207.011}

\bibitem[{{STSCI Development Team}(2012)}]{astrodrizzle}
{STSCI Development Team}. 2012, {DrizzlePac: HST image software}.
\newblock \doeprint{1212.011}

\bibitem[{{Sun} {et~al.}(2022){Sun}, {Leroy}, {Rosolowsky}, {Hughes}, {Schinnerer}, {Schruba}, {Koch}, {Blanc}, {Chiang}, {Groves}, {Liu}, {Meidt}, {Pan}, {Pety}, {Querejeta}, {Saito}, {Sandstrom}, {Sardone}, {Usero}, {Utomo}, {Williams}, {Barnes}, {Benincasa}, {Bigiel}, {Bolatto}, {Boquien}, {Chevance}, {Dale}, {Deger}, {Emsellem}, {Glover}, {Grasha}, {Henshaw}, {Klessen}, {Kreckel}, {Kruijssen}, {Ostriker}, \& {Thilker}}]{Sun22}
{Sun}, J., {Leroy}, A.~K., {Rosolowsky}, E., {et~al.} 2022, \aj, 164, 43, \dodoi{10.3847/1538-3881/ac74bd}

\bibitem[{{Tielens}(2010)}]{Tielens2010}
{Tielens}, A.~G.~G.~M. 2010, {The Physics and Chemistry of the Interstellar Medium}

\bibitem[{{Turner} {et~al.}(2021){Turner}, {Dale}, {Lee}, {Boquien}, {Chandar}, {Deger}, {Larson}, {Mok}, {Thilker}, {Ubeda}, {Whitmore}, {Belfiore}, {Bigiel}, {Blanc}, {Emsellem}, {Grasha}, {Groves}, {Klessen}, {Kreckel}, {Kruijssen}, {Leroy}, {Rosolowsky}, {Sanchez-Blazquez}, {Schinnerer}, {Schruba}, {Van Dyk}, \& {Williams}}]{turner21}
{Turner}, J.~A., {Dale}, D.~A., {Lee}, J.~C., {et~al.} 2021, arXiv e-prints, arXiv:2101.02134.
\newblock \doarXiv{2101.02134}

\bibitem[{{Walterbos} \& {Braun}(1994)}]{Walterbos94}
{Walterbos}, R. A.~M., \& {Braun}, R. 1994, \apj, 431, 156, \dodoi{10.1086/174475}

\bibitem[{{Watkins} {et~al.}(2023){Watkins}, {Barnes}, {Henny}, {Kim}, {Kreckel}, {Meidt}, {Klessen}, {Glover}, {Williams}, {Keller}, {Leroy}, {Rosolowsky}, {Lee}, {Anand}, {Belfiore}, {Bigiel}, {Blanc}, {Boquien}, {Cao}, {Chandar}, {Chen}, {Chevance}, {Congiu}, {Dale}, {Deger}, {Egorov}, {Emsellem}, {Faesi}, {Grasha}, {Groves}, {Hassani}, {Henshaw}, {Herrera}, {Hughes}, {Jeffreson}, {Jim{\'e}nez-Donaire}, {Koch}, {Kruijssen}, {Larson}, {Liu}, {Lopez}, {Pessa}, {Pety}, {Querejeta}, {Saito}, {Sandstrom}, {Scheuermann}, {Schinnerer}, {Sormani}, {Stuber}, {Thilker}, {Usero}, \& {Whitmore}}]{Watkins23}
{Watkins}, E.~J., {Barnes}, A.~T., {Henny}, K., {et~al.} 2023, \apjl, 944, L24, \dodoi{10.3847/2041-8213/aca6e4}

\bibitem[{{Whitmore} {et~al.}(2023){Whitmore}, {Chandar}, {Lee}, {Floyd}, {Deger}, {Lilly}, {Minsley}, {Thilker}, {Boquien}, {Dale}, {Henny}, {Scheuermann}, {Barnes}, {Bigiel}, {Emsellem}, {Glover}, {Grasha}, {Groves}, {Hannon}, {Klessen}, {Kreckel}, {Kruijssen}, {Larson}, {Leroy}, {Mok}, {Pan}, {Pinna}, {S{\'a}nchez-Bl{\'a}zquez}, {Schinnerer}, {Sormani}, {Watkins}, \& {Williams}}]{Whitmore23}
{Whitmore}, B.~C., {Chandar}, R., {Lee}, J.~C., {et~al.} 2023, \mnras, 520, 63, \dodoi{10.1093/mnras/stad098}

\bibitem[{{Williams} {et~al.}(2024){Williams}, {Lee}, {Larson}, {Leroy}, {Sandstrom}, {Schinnerer}, {Thilker}, {Belfiore}, {Egorov}, {Rosolowsky}, {Sutter}, {DePasquale}, {Pagan}, {Berger}, {Anand}, {Barnes}, {Bigiel}, {Boquien}, {Cao}, {Chastenet}, {Chevance}, {Chown}, {Dale}, {Deger}, {Eibensteiner}, {Emsellem}, {Faesi}, {Glover}, {Grasha}, {Hannon}, {Hassani}, {Henshaw}, {Jim{\'e}nez-Donaire}, {Kim}, {Klessen}, {Koch}, {Li}, {Liu}, {Meidt}, {M{\'e}ndez-Delgado}, {Murphy}, {Neumann}, {Neumann}, {Neumayer}, {Oakes}, {Pathak}, {Pety}, {Pinna}, {Querejeta}, {Ramambason}, {Romanelli}, {Sormani}, {Stuber}, {Sun}, {Teng}, {Usero}, {Watkins}, \& {Weinbeck}}]{Williams24}
{Williams}, T.~G., {Lee}, J.~C., {Larson}, K.~L., {et~al.} 2024, \apjs, 273, 13, \dodoi{10.3847/1538-4365/ad4be5}

\bibitem[{{Zhang} \& {Bloom}(2020)}]{Zhang20}
{Zhang}, K., \& {Bloom}, J.~S. 2020, \apj, 889, 24, \dodoi{10.3847/1538-4357/ab3fa6}

\end{thebibliography}


\appendix

\begin{figure*}[!htp]
    \centering
\includegraphics[width=\textwidth]{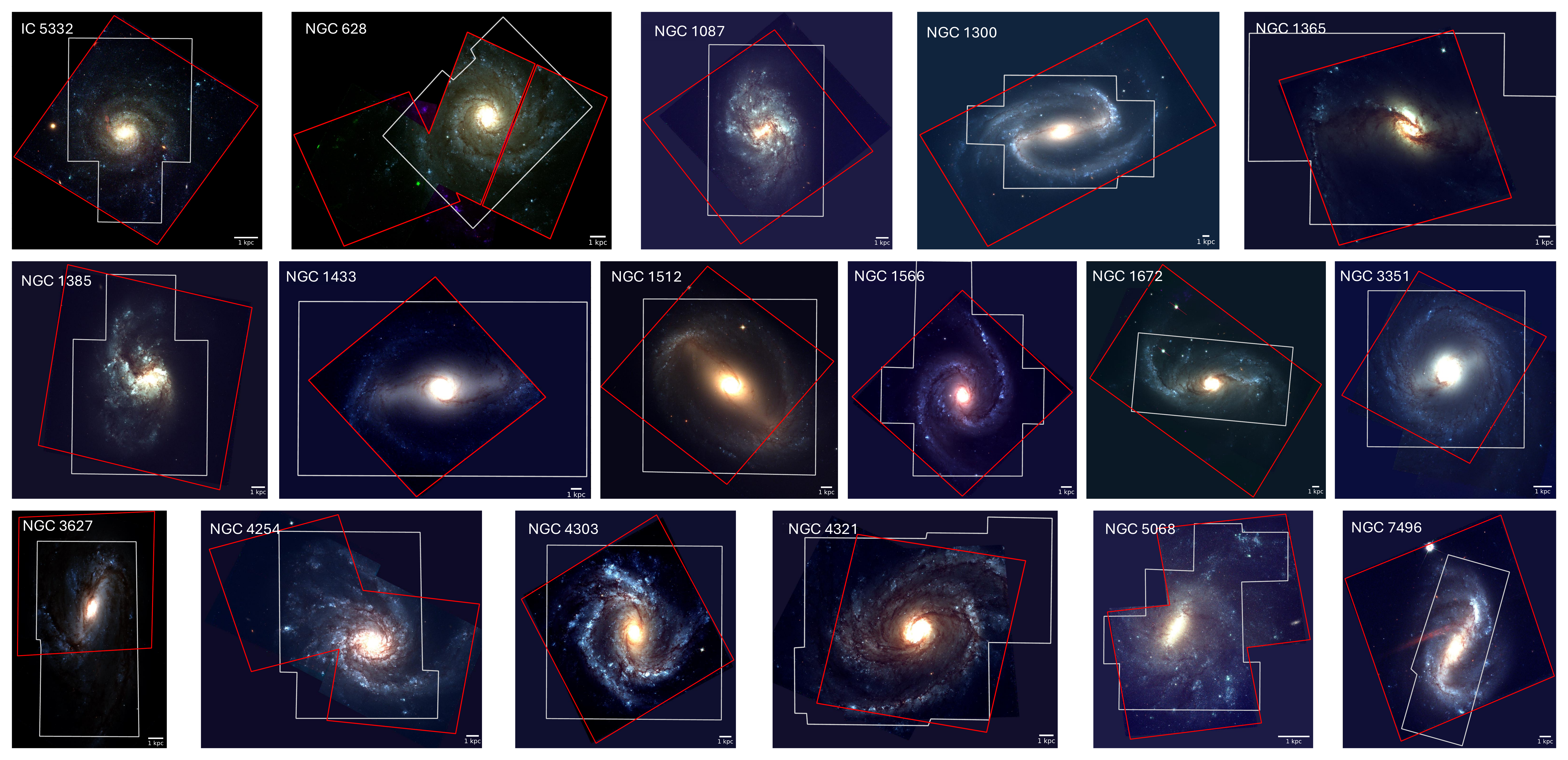}
    \caption{Footprint of the HST-H$\alpha$ images (red) and MUSE IFU observations (white) are shown on HST BVI color images for the 17 galaxies which currently have H$\alpha$ observations available.  A 1~kpc scale bar is shown in each panel.
    \label{fig:footprints} } 
\end{figure*}

\section{Additional Checks on Background and [NII] correction}
\label{appendix_fits}

In Section~\ref{sec_fluxan} we described our pipeline steps to produce high-resolution, calibrated HST-H$\alpha$ images.  MUSE IFU spectroscopy was used for  two purposes: to determine correct background levels for the HST-H$\alpha$ images and the amount of contamination from nearby [NII] emission lines which fall in the HST-H$\alpha$ passbands.  In this Appendix we present additional details to justify some of the detailed choices that were made.
In Figure~\ref{fig:footprints} we show footprints of the HST-H$\alpha$ images (red) and MUSE IFU observations (white), and 
note there is good overlap for most galaxies.

\subsection{Background and [NII] Level Determination From Different Fitting Parameters}

In Sections \ref{sec_fluxan} and \ref{subsec_fluxan4}, we determined accurate background levels for the HST-H$\alpha$ images and derived correction factors for the [NII] emission-line contamination in each galaxy by calibrating them to the MUSE spectroscopic observations. In both cases, we allowed the slope and intercept to be free parameters in the fits; however, we only applied the offset in the former case (background correction) and the slope in the latter ([NII] correction). In this section, we explore the impact on the derived correction factors of fixing the slope to unity or the offset to zero, following the same procedures outlined in Sections \ref{sec_fluxan} and \ref{subsec_fluxan4}.

Columns 2, 3, and 4 of Table~\ref{tab_bgcorr_niicorr_fixed} report the background level fits (the y-intercept) obtained when the slope is fixed to unity rather than treated as a free parameter. Column 5 of Table~\ref{tab_bgcorr_niicorr_fixed} lists the correction factors for the [NII] emission lines when the y-intercept is fixed to zero. We find that these values, derived under fixed-slope or fixed-intercept conditions are broadly consistent with those obtained when both slope and offset are allowed to vary freely.

\begin{table}
\centering

    \caption{Fitting results for the background correction in each of the HST filters (\S\,\ref{subsec_fluxan4}), and the correction factor for the [N\,II] $\lambda$6548\AA\ and [N\,II] $\lambda$6583\AA\ line contamination in the H\,$\alpha$ images (\S\,\ref{subsec_contsub2}). Here we have fixed the slope of the fit for the background correction to unity, and the intercept for the [N\,II] correction to zero (see Tab.\,\ref{tab_bgcorr_niicorr} where these are left as free parameters in our fitting).}
    \label{tab_bgcorr_niicorr_fixed}
    \begin{tabular}{ccccc}
\hline \hline
Galaxy & \multicolumn{3}{c}{Background Corrections} & [NII] Correction \\
& \multicolumn{3}{c}{[erg\,s$^{-1}$\,cm$^{-2}$\,\AA$^{-1}$\,arcsec$^{-2}$]} & \\
 & F555W & F65XN & F814W & Factor \\
\hline
    IC~5332 & -337.96 & 1838.07 & -227.22 & 1.10 \\
    NGC~628-C & -483.21 & -516.31 & -340.13 & 1.35 \\
    NGC~628-E & -149.63 & 1579.62 & -175.47 & 1.19 \\
    NGC~1087 & -272.16 & 828.05 & -181.57 & 1.19 \\
    NGC~1300 & -51.87 & -94.14 & -60.14 & 1.33 \\
    NGC~1365 & -882.44 & 438.51 & -723.04 & 1.27 \\
    NGC~1385 & -249.12 & 832.78 & -190.41 & 1.23 \\
    NGC~1433 & -620.67 & -663.53 & -544.69 & 1.43 \\
    NGC~1512 & -198.70 & 1435.68 & -128.10 & 1.03 \\
    NGC~1566 & -1022.72 & 1544.01 & -577.03 & 1.03 \\
    NGC~1672 & -250.62 & -363.09 & -171.77 & 1.39 \\
    NGC~2835-S & -505.99 & 715.02 & -369.64 & 1.00 \\
    NGC~3351 & -294.10 & 329.90 & -229.84 & 1.08 \\
    NGC~3627 & -1098.37 & -236.31 & -838.19 & 1.15 \\
    NGC~4254 & -77.69 & 93.99 & -229.99 & 0.97 \\
    NGC~4303 & -1140.68 & -566.77 & -900.57 & 0.97 \\
    NGC~4321 & -801.34 & 566.60 & -713.40 & 1.28 \\
    NGC~4535 & -648.72 & 605.50 & -531.96 & 1.17 \\
    NGC~5068 & -331.19 & -143.40 & -202.55 & 1.10 \\
    NGC~7496 & -172.27 & 549.13 & -134.12 & 1.25 \\
\hline
\end{tabular}
\end{table}

\subsection{HST-to-MUSE H$\alpha$ fluxes as a function of nebular properties}
\label{appendix_ratioplots}

\begin{figure*}[t]
    \centering
        \includegraphics[width=\textwidth]{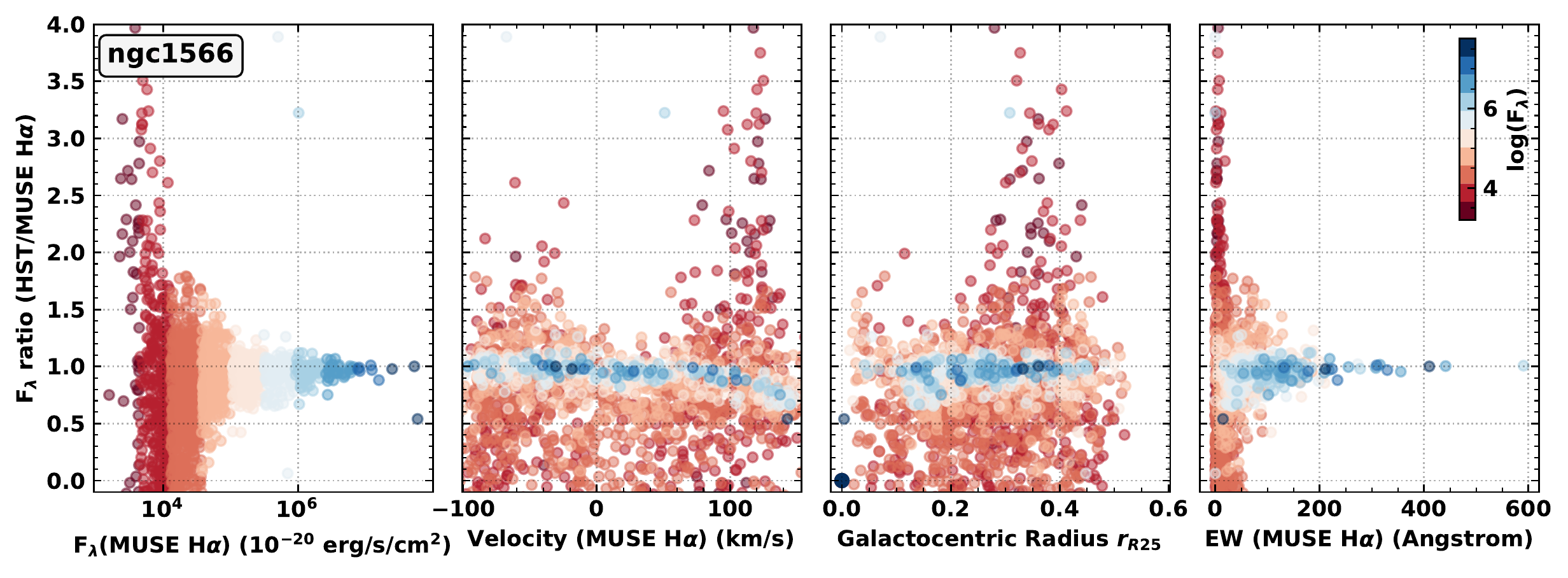} \\ \vspace{-7mm}
        \includegraphics[width=\textwidth]{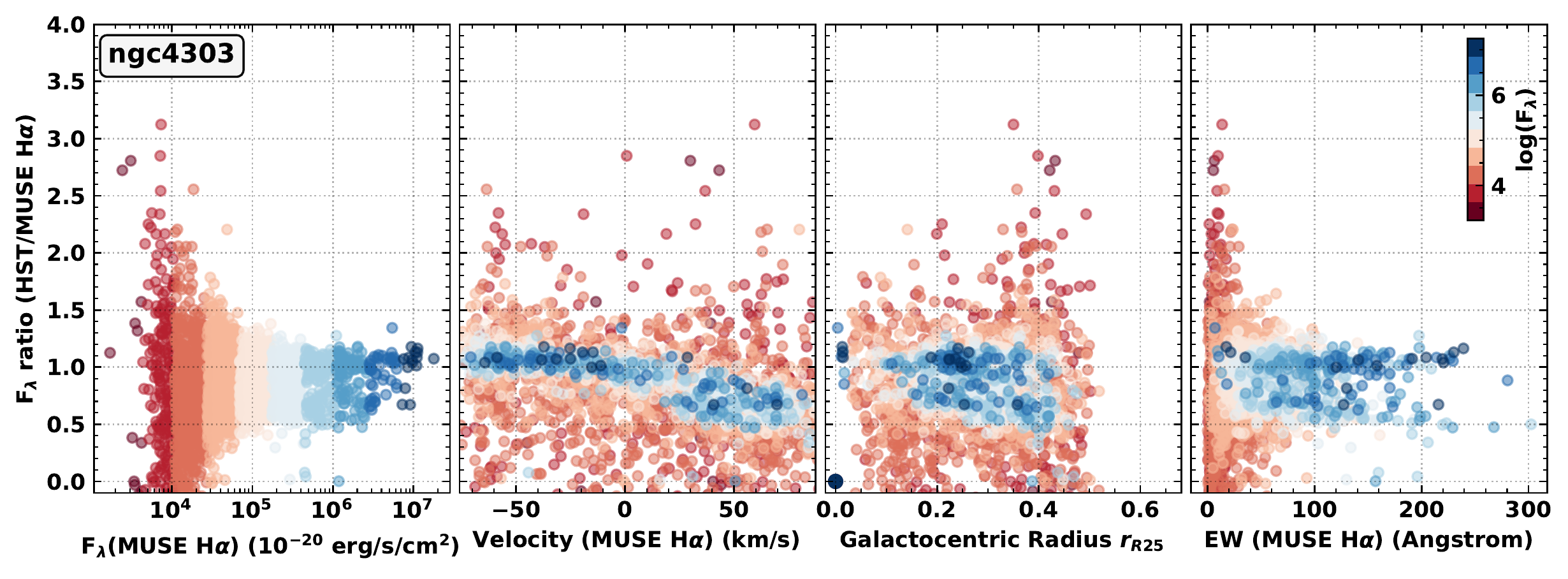} \\ \vspace{-3mm}
    \caption{\textbf{The ratio of the HST-to-MUSE H$\alpha$ fluxes checked against other nebula properties for NGC~1566 and NGC~4303.} We show (from left to right) the H$\alpha$ flux ratio versus (left panel) the MUSE H$\alpha$ flux, (second panel) MUSE H$\alpha$ velocity, (third panel) galactocentric radius, and (right panel) the H$\alpha$ equivalent width measured directly from the MUSE spectra. There is no obvious trend in the flux ratio with H$\alpha$ velocity or galactocentric distance (middle two panels), although for NGC~4303, there appears to be a decline in the HST-to-MUSE flux ratio at higher velocities ($>$\,100\,km~s$^{-1}$), likely due to the filter response.}
    \label{fig_appendix1}
\end{figure*}

In Figure~12, for NGC~628c we showed the ratio of HST-to-MUSE H$\alpha$ fluxes versus the MUSE H$\alpha$ flux, MUSE H$\alpha$ velocity, galactocentric radius, and the H$\alpha$ equivalent width measured directly from the MUSE spectra. 
There is some range in the results for the galaxies in our sample, so in Figure~\ref{fig_appendix1}, we present the exact same measurements for NGC~1566 (top panels) and NGC~4303 (bottom panels). 
These plots allow a deeper understanding of any systematic variations in the HST-to-MUSE flux ratios, as well as the reliability of cross-calibration for H$\alpha$ measurements across different galactic conditions.

Unlike the results shown in the main text for NGC~628 (Fig.~\ref{fig_Ha_ratio2}), here we observe a trend of decreasing HST-to-MUSE flux ratios at higher velocities ($>$\,50\,km~s$^{-1}$), which is particularly pronounced for NGC~4303 (also see $>$\,100\,km~s$^{-1}$ for NGC\,1566). 
This effect is likely due to the position of NGC~4303 at the edge of the F658N filter response curve (see Fig.\,\ref{fig:filter}), which could lead to a reduction in the effective H$\alpha$ transmission at higher velocities relative to MUSE. 
In contrast, NGC~628 does not show this trend in Figure~\ref{fig_Ha_ratio2}, supporting the hypothesis that proximity to the filter band edge influences flux measurements in cases with high nebular velocities.
To correct for this, velocity dependent flux correction will be included in a future version of the pipeline.

\subsection{F555W filter versus F550M and F547M for continuum subtraction}
\label{appendix_555filters}

The F555W filter was used (along with F814W) to estimate and subtract the continuum level in the narrow-band images.  This filter may suffer contamination from strong emission lines such as [OIII] and potentially introduce residuals in continuum subtraction that affect the derived flux maps when compared with the narrower F550M or F547M filters.
We tested the impact of using the F550M or F547M filters rather than the F555W filter
by using data available for NGC\,1672 and NGC\,3351.
We ran the full pipeline, which includes the background subtraction and [NII] correction steps (using the F550M or F547M filter transmission curves applied to the MUSE cubes), as well as using the filters for the continuum subtraction. 

\begin{figure*}[t]
    \centering
        \includegraphics[width=\textwidth]{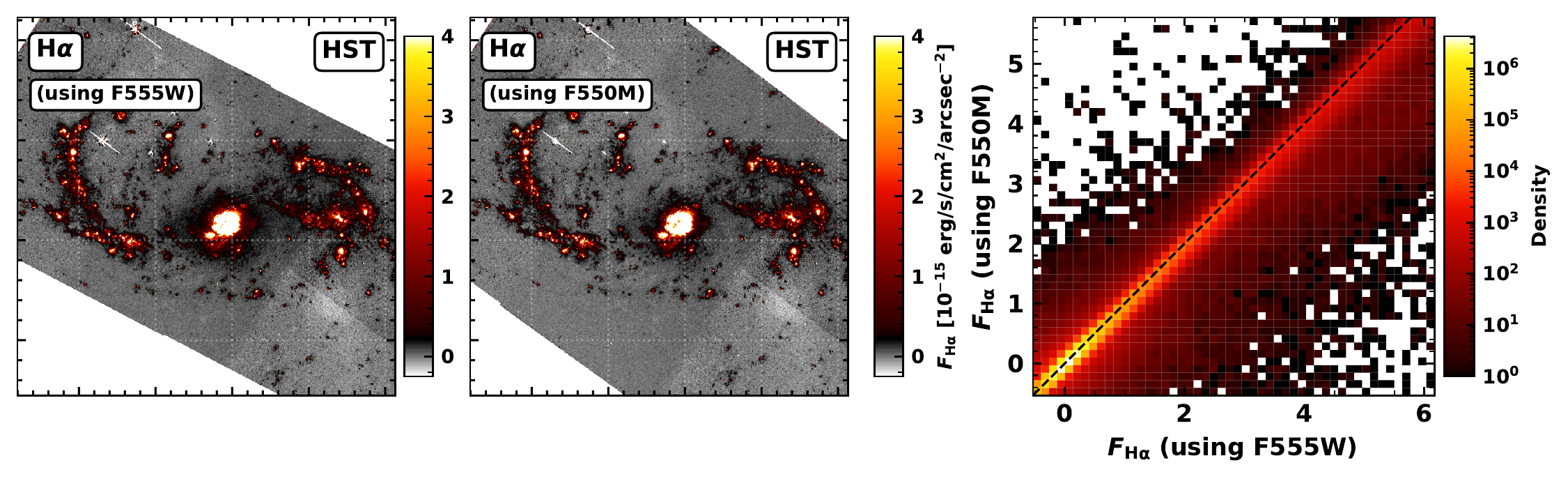} \\ 
        \includegraphics[width=\textwidth]{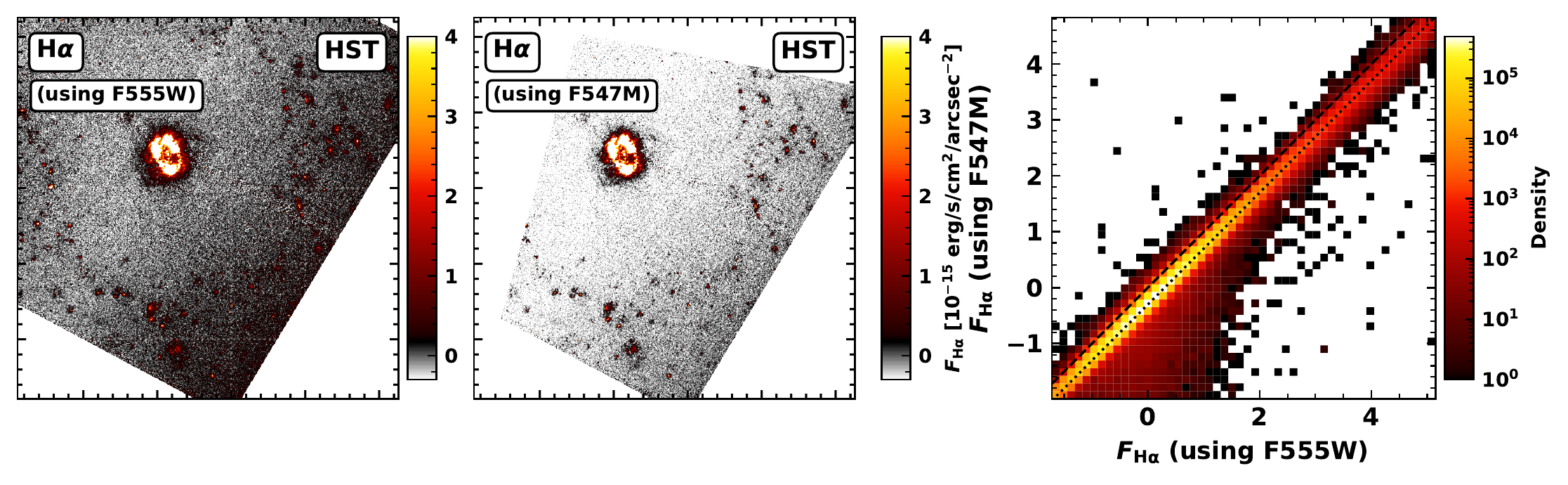}   
    \caption{\textbf{Difference between using the F555W versus F550M and F547M filters for continuum subtraction.}
    The top row shows the difference between using the F555W (left) and F550M (middle) filters for continuum subtraction in NGC\,1672, while the lower row shows the difference between F555W and F547M for NGC\,3351.
    The right panels show 2D histogram density plots of the \Ha\ flux at each pixel for both cases. Overlaid is a dashed line showing the $y=x$ relation, and the dotted line in the bottom panel shows an offset of $y=x-0.3$.}
    \label{fig_appendix2}
\end{figure*}

Figure\,\ref{fig_appendix2} shows the resulting maps and a comparison of the per-pixel fluxes.
We observe that in NGC\,1672 the F555W filter appears to have some minor residual continuum in the image when compared to the F550M filter subtraction, which appears predominantly towards the center of the galaxy. 
This is consistent with expectations that [OIII] emission might contaminate the F555W filter more significantly in regions with strong ionizing sources. 
Nonetheless, overall we see a very close one-to-one match of the fluxes across the galaxy (see top-right panel of Fig.\,\ref{fig_appendix2}). 
In NGC\,3351 we see that there is a slight shift to lower values for the F547M filter subtraction when compared to the F555W subtraction (see dotted line in Fig.\,\ref{fig_appendix2}), though this appears to be an effect of poor background subtraction in the F547M filter rather than to line contamination, which would predominantly effect smaller scales. 
Overall, we found that switching to the F547M or F550M filters makes little difference to the final flux-calibrated, continuum-subtracted images, indicating that line contamination does not significantly propagate into the final products for these galaxies.

\end{document}